\documentclass[review]{elsarticle}
\makeatletter
\def\ps@pprintTitle{%
 \let\@oddhead\@empty
 \let\@evenhead\@empty
 \def\@oddfoot{\centerline{\thepage}}%
 \let\@evenfoot\@oddfoot}
\makeatother

\usepackage{lineno,hyperref}
\usepackage{subfig}
\usepackage{xcolor}
\usepackage{booktabs}
\usepackage{amsmath, amsthm, amssymb, amsfonts}
\usepackage{caption}
\usepackage[T1]{fontenc}
\usepackage{float}

\newlength\m
\newlength\z

\modulolinenumbers[5]

\newtheorem{definition}{Definition}
\journal{International Journal of Forecasting}

\begin{document}

\begin{frontmatter}

\title{Anchor Attention for Hybrid Crowd Forecasts Aggregation}

%\author{Anonymous}
\author[isi]{Yuzhong Huang\corref{mycorrespondingauthor}}
\ead{yuzhongh@isi.edu}
\author[isi]{Andr\'{e}s Abeliuk}
\ead{aabeliuk@isi.edu}
\author[isi]{Fred Morstatter}
\ead{fredmors@isi.edu}
\author[pytho]{Pavel Atanasov}
\ead{pavel@pytho.io}
\author[isi]{Aram Galstyan}
\ead{galstyan@isi.edu}
\address[isi]{Information Sciences Institute, University of Southern California}
\address[pytho]{Pytho, LLC.}

\cortext[mycorrespondingauthor]{Corresponding author}

\begin{abstract}
%Most forecast aggregation methods are based on linear average.\cite{atanasov2017distilling}. 
In a crowd forecasting system, ``aggregation'' is an algorithm that returns aggregated probabilities for each question based on the probabilities provided per question by each individual in the crowd. Various aggregation methods \cite{atanasov2017distilling, l2reg} have been proposed, but simple strategies like linear averaging or selecting the best-performing individual remain competitive. With the recent advance in neural networks, we model forecasts aggregation as a machine translation task, that translates from a sequence of individual forecasts into aggregated forecasts, based on proposed Anchor Attention between questions and forecasters. We evaluate our approach using data collected on our forecasting platform and publicly available Good Judgement Project dataset~\cite{GJP}, and show that our method outperforms current state-of-the-art aggregation approaches by learning a good representation of forecaster and question.
\end{abstract}

\begin{keyword}
Aggregation\sep Crowd Sourcing\sep Embedding\sep Attention Model
\end{keyword}

\end{frontmatter}

%\linenumbers

\section{Introduction}
Forecasting the outcome of geopolitical events is a notoriously difficult problem, where even experts in the domain area relevant to the forecasting problem fail to significantly outperform simple extrapolation algorithms~\cite{tetlock2017expert}. One way to address this challenge is to pose the problem to a crowd of forecasters, i.e., to use a crowdsourcing approach.

Recent attention has been devoted to producing geopolitical forecasting systems using crowdsourcing methods. These efforts were propelled by the Intelligence Advanced Research Project Activity (IARPA), who in 2011 launched a research program called Aggregate Contingent Estimation (ACE) to enhance the accuracy and precision of crowdsourced geopolitical forecasting systems.\footnote{\url{https://www.iarpa.gov/index.php/research-programs/ace}} The wisdom of the crowd approach was able to generate accurate forecasts across a wide range of forecasting problems, relative to professional intelligence analysts with access to classified information ~\cite{tetlock2016superforecasting}. 

To produce a high-quality final forecast, an essential step is to aggregate these crowd forecasts. Each forecaster could have different background and expertise, and some "forecasters" could be algorithm models based on collected data. Combining these sources should lead to an \emph{aggregated forecast} that is more accurate than those produced by either source independently. Although assessing the quality of each source is important to understanding the best way to combine them, the strategy for combining these sources of input is not clear. Previous work~\cite{atanasov2017distilling,satopaa2014combining} has identified linear combinations among forecaster estimates. However, such an approach may under-fit available data, since it assigns a single global weight to each forecaster in a variety of forecasting problems at any moment.

In this paper, we focus on developing a new approach to aggregate human crowd and machine-generated forecasts using the latest advancements in neural networks. The key insight is that our proposed Anchor Attention model learns a representation that can infer the best aggregation weight for a forecast made by a forecaster for a given question. The weight is conditioned on the question, forecaster, and time. It is more flexible than a single weight per forecaster as used in previous methods.

Specifically, our contributions are the following: \begin{enumerate}
  \item We formulate forecast aggregation as a neural machine translation task.
  \item We analyze the issue of self-attention \cite{Vaswani2017AttentionIA} in forecasts aggregation setup and propose Anchor Attention model that is tailored for this task, and robust to low-quality forecasts.
  \item We evaluate our model using data from our platform and Good Judgement Project dataset and show it outperforms baselines~\cite{atanasov2017distilling, l2reg}.
\end{enumerate}
	
\section{Related Work}
Approaches for weighting and combining probabilistic judgments have attracted sustained interest in the forecasting literature~\cite{armstrong1989combining,clemen1989combining, jose2008simple, winkler2019probability,lichtendahl2020some}. A special case relevant to the problem described here is when forecasts arrive at different times~\cite{brown1991forecast}. Although linear aggregation methods are the most popular approach for combining probability forecasts, R. Ranjan \cite{ranjan2010combining} proved theoretically that this approach is sub-optimal for fixed weights and calibrated predictions. Thus, a non-linear re-calibration or extremization is widely used to improve performance~\cite{baron2014extremizing}. This has motivated other methods for aggregating probabilities in a non-linear way. V.A. Satop{\"a}{\"a}~\cite{satopaa2014combining} introduced a model-based aggregator that maps the probabilities into the log-odds space, allowing modeling the probabilities using the normal distribution. This approach has been extended in \cite{satopaa2014probability} to exploit the temporal correlation between forecasts, similar to standard time-series approaches like auto-regressive integrated moving average (ARIMA), where future forecasts are a linear combination of past forecasts. 
	
Pertaining to neural network-based aggregation methods, to the best of our knowledge, there are two relevant papers.  
A. Gaunt \cite{gaunt2016training}  implemented a network of three fully-connected layers, tailored for tasks such as data annotation or labeling, to aggregate crowd opinions. This model is not designed to learn the temporal relation and historical performance of the forecasters, and requires every participant to respond to every question.
	G. Nebbione \cite{crowd_ranking} proposed a method capable of learning from historical performance, but it focuses on ranking individual forecasts, rather than aggregating and making a better forecast.

Sequence-to-sequence is a general framework widely used in NLP communities. It has been used in tasks like machine translation~\cite{NIPS2014_a14ac55a}, voice recognition~\cite{Zhang2019SequencetoSequenceAM}, text summarization~\cite{10.1145/3419106}. Architecture side, Elman and Jordan proposed recurrent neural network \cite{Elman90findingstructure, JORDAN1997471}, the first model computationally practical model to learn structures from time. S. Hochreiter proposes LSTM model \cite{lstm}  which greatly extends its memory capacity. A. Vaswani removed the recurrent component and proposed self-attention model \cite{Vaswani2017AttentionIA}, which allows fast parallel computations for all intermediate states. This series of work inspires us to apply sequence-to-sequence model on forecasts aggregation. However, with our experiments, existing models are not ideal for this task. We will elaborate it in Section \ref{sec:self}.

	\section{Aggregation Setup}
    The general framework of crowdsourced platforms for forecast generation consists of two main components. First, acquire information from human experts regarding predictions about future events. Techniques for eliciting crowd information range from prediction markets~\cite{wolfers2004prediction,yeh2006using} to prediction polls~\cite{mellers2015identifying,tetlock2016superforecasting}.
	Second, aggregation algorithms are developed to generate a prediction about future events based on all the information gathered from the crowd.
    
	%We will compare our methods to those proposed by P. Atanasov ~\cite{atanasov2017distilling} and D. Cross \cite{l2reg}, which are described in detail in Section~\ref{sec:baseline_methods}.
	
	%Our work studies geopolitical event forecasts created on our forecasting platform: Synergistic Anticipation of Geopolitical Events (SAGE)\cite{sage}\footnote{\url{https://sage-platform.isi.edu/}}. %, Synergistic Anticipation of Geopolitical Events (SAGE).\footnote{URL anonymized for blind review.} 
    In this work, we apply our proposed aggregation method to both Good Judgment Project (GJP) dataset \cite{GJP} and our hybrid forecasting platform (HFC).
    %\footnote{URL anonymized for blind review.}
    These datasets contain a set of questions pertain to categorical or ordinal geopolitical events, and each contains 2 to 5 non-overlapping answer options. Each question remains open (available for forecasting) for a predetermined amount of time. Forecasters can generate forecasts while the question is open. The forecasting system will make a daily aggregated forecast during the forecast period. On the resolution date, the correct answer will be published with a link to the source that determines the correct answer. The performance of the forecasting system will be evaluated based on their generated daily aggregated forecast, using mean daily Brier Score~\cite{brier1950verification} as described in Section \ref{sec:brier}.
    
    \subsection{Good Judgement Project Dataset (GJP)}
    The Good Judgement Project~\cite{ungar2012the} is a large-scale study to find the best way to use a set of experts to estimate the probability of a future event. The released dataset \cite{GJP}, contains 4 years of forecasting data and more than 600 forecasting problems, is a valuable resource for aggregation researcher, and inspire future research projects like our hybrid forecasting platform.
    
    \subsection{Our Hybrid Forecasting Platform (HFC)}
\begin{figure}[t]
	\centering
	\subfloat[][What will be the long-term interest rate for Hungary in July 2018? \\\textit{Correct Answer:} More than 2.8]{    \includegraphics[width=0.44\textwidth]{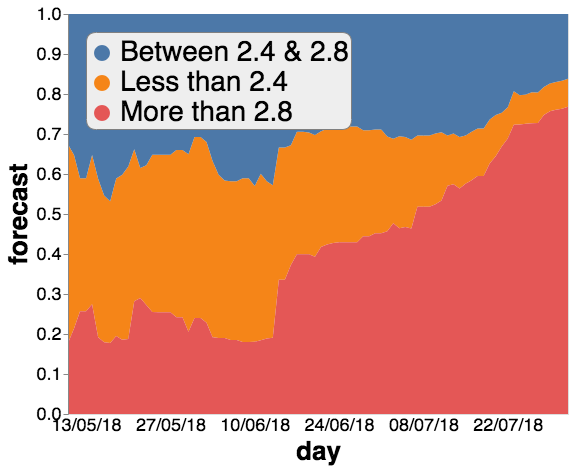}\label{fig:example1}
	}
	\hfill
	\subfloat[][Will UK's Prime Minister Theresa May vacate office between 6 December 2018 and 15 January 2019?\\ \textit{Correct Answer:} No]{      \includegraphics[width=0.44\textwidth]{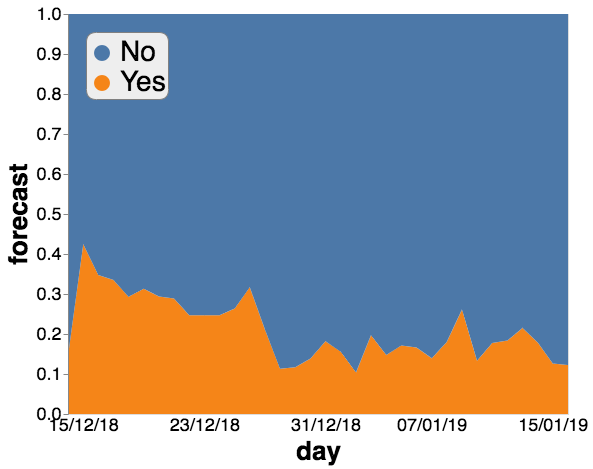}\label{fig:example2}
	}
	\caption{Daily average forecasts for two sample questions. }
	\label{fig:example}
\end{figure}

    Our hybrid forecasting platform mostly follows the same setup as the good judgment project, but adds machine models for questions that have available historical data.  Both humans and machines enter their forecasts by assigning a probability to a set of predefined answer options. Figure \ref{fig:example} illustrates two examples showing the daily average forecast for each question. Notice that as the resolution date comes nearer, forecasters have more and new information available; thus the consensus shifts towards the correct answer, for both questions. For this reason, our methods that dynamically assign weights for each forecaster are more flexible and have the potentials for better performance.

    \subsubsection{Human Forecasters}
    The study's human forecasters were recruited through the Amazon Mechanical Turk platform and advertisements on blogs and periodicals related to geopolitical forecasting. Recruitment took place before the study began. Forecasters were free to provide forecasts to any number of questions. To maintain user engagement, new questions were added to the platform each week. The questions were active from 3 days to 184 days. Forecasters were allowed and encouraged to update their predictions as they received new information over time. Forecasters are paid \$16 for two hours of work and are invited back each week to submit forecasts on new questions, and to update existing ones. 
    %The dataset consists of human forecasts and the forecasts generated by machine models on the same questions. Anonymized forecast data and questions will be included as supplementary materials.

	Overall, throughout our hybrid forecasting project, we had $2,240$ participants producing $98,258$ forecasts on $375$ questions. Table~\ref{tbl:summary} shows a summary of the data. There was a wide variance in the number of forecasts that a given question received. This was due to the question's difficulty, primarily stemming from the amount of research required to generate a forecast.
	
	\subsubsection{Machine Models}
	Some types of questions have historical data available. For example, financial questions usually have historical data provided by a stock exchange. For instance, question \emph{What will be the long-term interest rate for Hungary in July 2018}, are backed by reliable data and there is a good chance that machine models would work well.
    
    Given that questions covered a broad set of geopolitical topics and different data sources, we used a general approach to extrapolate time series to produce machine forecasts from the historical data relevant to each question. Machine models are first fit to historical time series, then extrapolate the time series to predict mean and variance. Assume the target value follows a normal distribution, we convert the estimated probability density distribution into probabilities in each answer option. We used three types of machine models: 1) AutoRegressive Integrated Moving Average (ARIMA)~\cite{arima}; 2) M4-Meta \cite{m4}; and 3) Arithmetic Random Walk (RW).
    
	\section{Methodology}
	\subsection{Neural Machine Translation}
	To better understand the need for our proposed Anchor Attention, we are doing a brief overview of neural machine translation and attention layer. The attention model is initially introduced by Dzmitry Bahdanau \cite{joint_align} in 2014, to be used in neural machine translation. There are 4 steps to neural machine translation. In this section we will illustrate these steps by translating the English sentence “have a good day” to German. See Fig. 2 for an illustration.
	
	The first step is word embedding. We need a word embedding table $\mathbf{E}$ in the shape of $|V| \times |D|$, where $|V|$ is vocabulary size, the number of unique words in the training corpus, usually in the range of $10^3$ to $10^5$,  and $|D|$ is embedding dimension, usually in the range of 10 to 300. The word embedding table is a model parameter that will be updated in the training process. There are 4 words in the source sentence, and their respective embedding vector $\mathbf{x_1}$ to $\mathbf{x_4}$ was selected from the embedding table.
	
	The second step is encoding, which computes a sentence representation $\mathbf{s}$ of these words. In machine translation, a recurrent encoder is widely used. Various recurrent encoder has been proposed~\cite{lstm,gru,elstm}. Recurrent encoder could be summarized as a randomly initialized internal state vector $\mathbf{s}$, and a set of weight matrices $\mathbf{W}$ that defines how internal state would update given an input. So as $\mathbf{x_1}$ to $\mathbf{x_4}$ feed into recurrent encoder, its internal state is updated from $\mathbf{s_1}$ to $\mathbf{s_4}$. Then we could take $\mathbf{s_4}$ as the sentence embedding $\mathbf{s}$.
	
	The third step is decoding, which is the inverse of the encoding step. The recurrent decoder has the same structure as the recurrent encoder with a different set of weight $\mathbf{W'}$, but its initial internal state vector is set to $\mathbf{s}$, the final encoder state. The first input to the decoder is a special end of sequence token $<$eos$>$, and its internal state is updated to $\mathbf{t_1}$.

	The fourth step is word prediction. It's formulated as a $|V'|$ way classification problem that using a fully connected layer with weight $W^{out}$ to predict target token $\hat{y_i}$ from all possible tokens in target language, $|V'|$ is the vocabulary size of target language. The newly translated word will be embedded using target language embedding table $\mathbf{E'}$, and fed into the decoder as the next input. This loop continues until the special token $<$eos$>$ is produced.

    These 4 steps illustrate the workflow of machine translation. There are 5 groups of parameters: $\mathbf{E, E', W, W', W^{out}}$, that would be learned in the training stage. In summary, source sentence $\{\mathbf{x_i}\}$ will first be encoded using a source language encoder into a language-independent sentence representation $\mathbf{s}$, then be decoded by target language decoder into target language sentence $\{\mathbf{y_i}\}$. The sequence length for the source and target language could be different, which makes it possible to translate between very different languages. This architecture is very flexible that it has been extended to various tasks, like video caption, which is formulated as a translating from a sequence of images to a sequence of words.

	\begin{figure}[!tbp]
      \centering
      \begin{minipage}[b]{0.49\textwidth}
        \includegraphics[width=\textwidth]{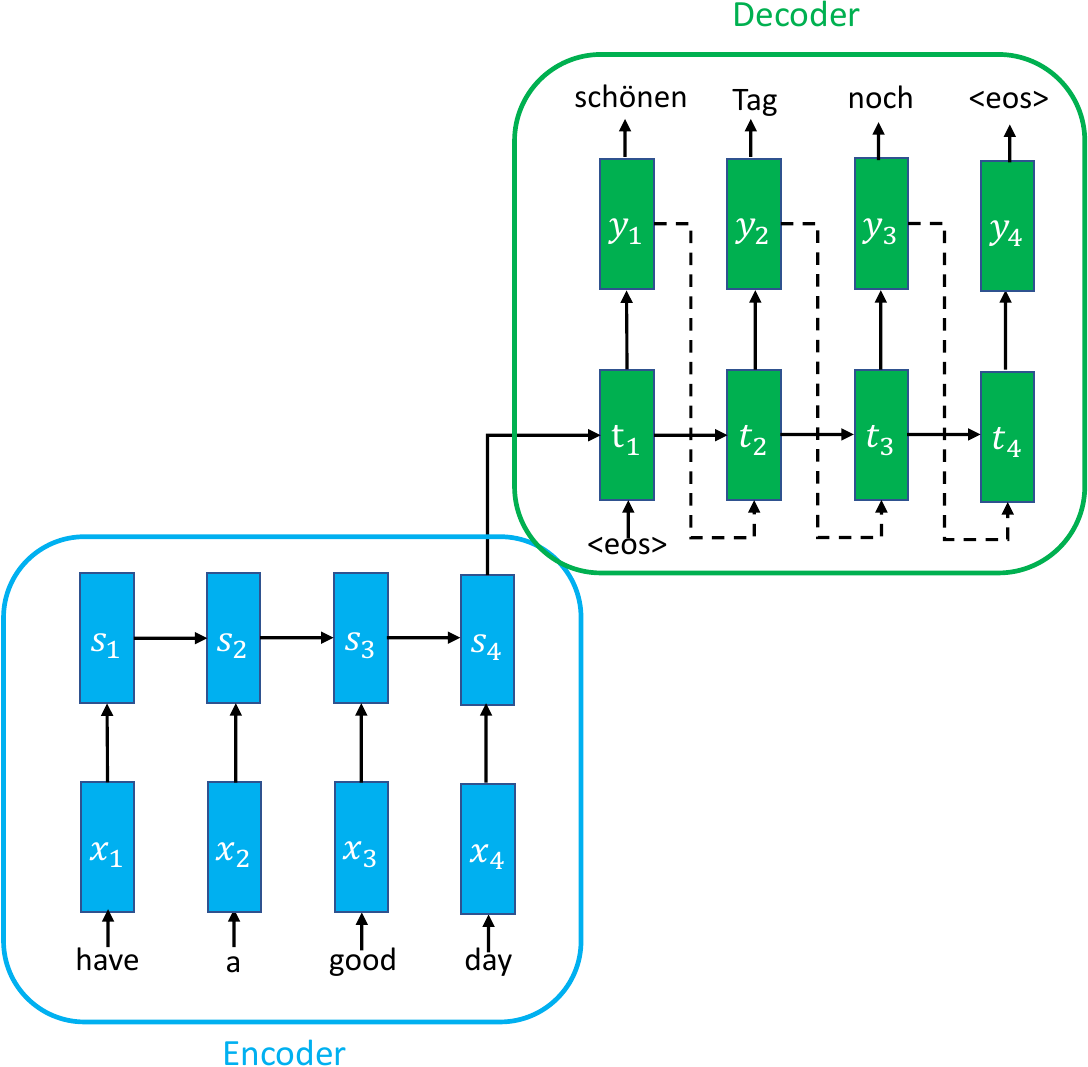}
        \label{fig:lstm_att_layout}
        \caption{Encoder-Decoder Architecture}
      \end{minipage}
      \hfill
      \begin{minipage}[b]{0.49\textwidth}
        \includegraphics[width=\textwidth]{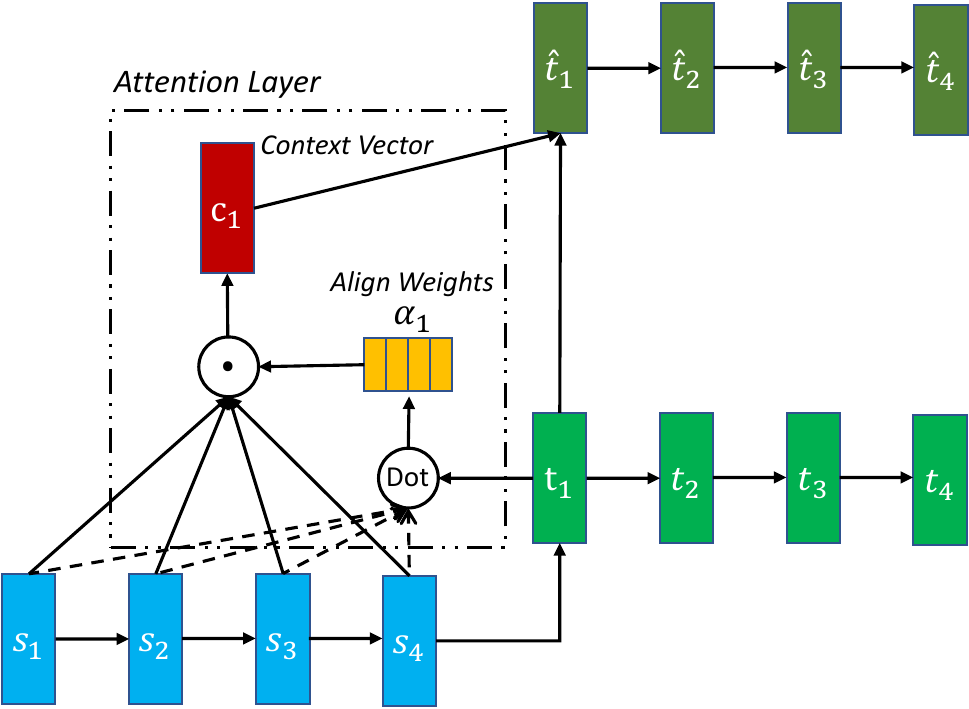}
        \caption{Attention Layer}
      \end{minipage}
    \end{figure}

    \subsection{Attention Layer}
    In the original neural machine translation architecture, all the information from the source language must be encoded into sentence representation $\mathbf{s}$. If the source sequence becomes longer, this quickly becomes a bottleneck, as it's against information theory to encode a very long (infinite) sequence into a fixed-length recurrent state vector. Therefore, Dzmitry Bahdanau \cite{joint_align} introduced the attention layer in 2014, as illustrated in Fig. 3.
	
	The introduction of attention layer changes the scope of word prediction step. In non-attention model, only the decoder state $\mathbf{t_i}$ is used for word prediction. In attention model, the decoder state $\mathbf{t_i}$ is enhanced with context vector $\mathbf{c_i}$, which is a weighted average of all previous encoder state, formulates a contextualized decoder state $\mathbf{\hat{t}_i}$ to be used in word prediction. There are various choices of how $\mathbf{\hat{t}_i}$ is computed, a simple approach is: \begin{align}
	    \alpha_{i}(j) = \frac{exp(\mathbf{t_i \cdot s_j})}{\sum_{j'} exp(\mathbf{t_i \cdot s_j'} )}\\
	    \mathbf{c_i} = \sum_{j} \alpha_{i}(j) \cdot \mathbf{s_j}\\
	    \mathbf{\hat{t}_i} = tanh(\mathbf{W^c [c_i; h_i]})
	\end{align}

    There are three steps to compute contextualized decoder state $\mathbf{\hat{t}_i}$.
    \begin{enumerate}
        \item Compute the alignment weights vector. It's a vector having the same length as the input. Each element in the alignment vector is a dot product between the current decoder state $\mathbf{t_i}$ and corresponding encoder state $\mathbf{s_j}$, and normalized by softmax. 
        \item Compute context vector $\mathbf{c_i}$ as a weighted average of $\mathbf{s_j}$ and corresponding alignment weight $\alpha_i(j)$.
        \item Concatenate $\mathbf{c_i}$ and $\mathbf{h_i}$, and apply a projection weight $\mathbf{W^c}$ to get the contextualized decoder state $\mathbf{\hat{t}_i}$. The projection weight $\mathbf{W^c}$ needs to be learned during training.
    \end{enumerate}

	The attention layer greatly improves model's memory capacity, that instead of encoding all the information into $\mathbf{s}$, the model could look into different parts of the input sequence when generating the output sequence.
	
	\subsection{Transformer Model}
	The attention layer is initially proposed as an augmentation to the recurrent unit, but the above derivation shows the context vector $\mathbf{c_i}$ also has memory capacity, it could replace the recurrent unit and be used independently. In 2017, Ashish Vaswani proposed Transformer \cite{Vaswani2017AttentionIA}, which uses a stack of self-attention layers without any recurrent unit. The core element is scaled dot-product attention, illustrated in Fig. \ref{fig:scaleddot}
	
	\begin{minipage}{0.45\textwidth}

    \centering
    \includegraphics[width=0.5\linewidth]{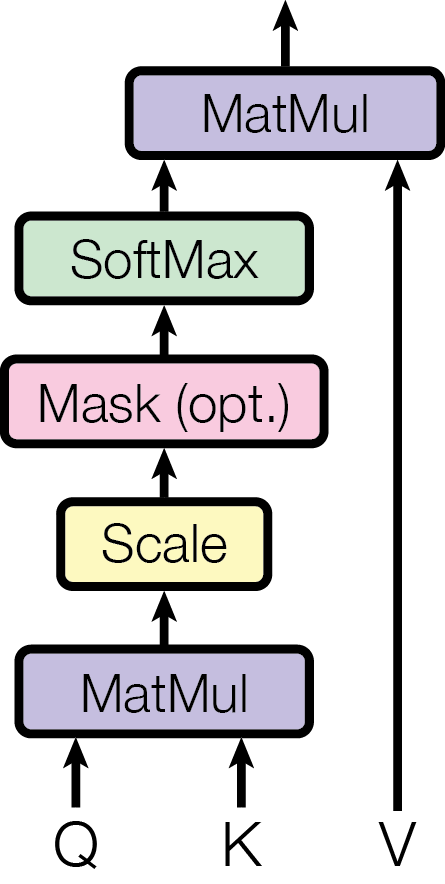}
    \captionof{figure}{Scaled Dot Product Attention}
    \label{fig:scaleddot}

    \end{minipage}%%% to prevent a space
    \begin{minipage}{0.55\textwidth}
    \scriptsize
	\begin{equation}
	\mathbf{ q_i = x_i \cdot W^Q, k_i = X_i \cdot W^K, v_i = x_i \cdot W^V}
	\end{equation}
	\begin{equation}
	\alpha_{ij} = align(\mathbf{q_i, k_j}) = \frac{exp(\mathbf{q_i \cdot k_j^T} / \sqrt{d_k})}{\sum_{j'} exp(\mathbf{q_i \cdot k_{j'}^T} / \sqrt{d_k})}
	\end{equation}
	\begin{equation}
	h_i = (\mathbf{v_1, v_2, \cdots, v_i}) \cdot (\alpha_{i1}, \alpha_{i2}, \cdots, \alpha_{ii})^T,
	\end{equation}
    \null
    \par\xdef\tpd{\the\prevdepth}
    \end{minipage}

	\vspace{1em}
	The transformer model introduced three trainable weight matrices: query, key, and value matrices. These matrices projects an input $\mathbf{x_i}$ into three vectors $\mathbf{q_i, k_i, v_i}$. As shown in Eq.(5), query vector $\mathbf{q_i}$ will be used to compute out-going attention score for current input, key vector $\mathbf{k_i}$ will be used to compute attention score from other inputs, value vector $\mathbf{v_i}$ will be used to compute final output $\mathbf{h_i}$. Using projected vector $\mathbf{q, k, v}$ instead of $\mathbf{x}$ directly, allows them to capture different information for their different needs, while reducing their dimension and lowering computing complexity. Output state $\mathbf{h_i}$ is analogy to $\mathbf{\hat{t}_i}$ in Eq. (3), that represents information in all inputs until time step $i$.
	
	The memory in a self-attention layer has an advantage over a recurrent unit, that it will not decay over a long sequence. According to Eq.(5), the attention score $\alpha_{i, j}$ will not change no matter $i, j$ are close or far away. While in a recurrent unit, the hidden state is computed step by step, the impact of an input would decay over time. Also, since transformer model removes the temporal dependency between recurrent states, and all states $\mathbf{h_i}$ could be computed in parallel, enabling much faster training and inference. The transformer model inspired a line of future works like BERT~\cite{bert}, GPT~\cite{gpt}, T5~\cite{t5}, they have similar architecture but much larger training corpus, achieving impressive results on various NLP tasks.

    \subsection{Using Attention in Aggregation}\label{sec:self}
    As neural machine translation architecture is flexible to translate any language pairs, we can treat individual forecasts as the source language, the aggregated forecasts as the target language, and use neural machine translation methods to learn to translate them. Consider a multiple-choice question with $k$ answer options. An individual forecast could be represented as $\mathbf{x_i \in R^k}$. Denote $\mathbf{x_t}$ as the last forecast on a day, to compute aggregated forecast at the end of that day, we could use output state $\mathbf{h_t}$ projected with output weight $W^O$ to produce the aggregated forecast $\mathbf{y_t \in R^k}$. We will elaborate on the setup in Sec. 4.6.
    
    This approach is similar to the setup in neural machine translation, but it has a drawback rooted in the nature of forecasting data. From Eq. (5), $\mathbf{h_i}$ is most influenced by $\mathbf{x_i}$ (the last input), because its projection $\mathbf{q_i}$ will be used as the query vector to compute the alignment score with all previous forecasts. For language tasks, this is not a problem, as the last token usually contains useful information, like ``?'' implies this sentence is a question. In the task of forecast aggregation, this means the aggregation weight for each forecast is determined by their similarity to the last forecast, which is \emph{not} desired. Any forecast can be the last forecast at the moment of making an aggregated forecast. The output of the aggregation system should not be oversensitive to the last forecast.

    \subsection{Anchor Attention}
    Based on the above discussion and the issue of using self-attention on the forecasts aggregation task, we proposed Anchor Attention. Anchor attention uses an anchor vector $\mathbf{a}$, which is independent of input sequence $\mathbf{x}$, to replace the query vector $\mathbf{q}$. Here, $\mathbf{a}$ is the sentence embedding of question text that captures the semantics of the question, $\mathbf{e_{text}}$ is question text embedding. Replacing Eq. (5), the alignment score is calculated as:
    \begin{align}
        \mathbf{a} &= tanh(\mathbf{e_{text} \cdot W^a}) \\
        \alpha_{ij} &= align(\mathbf{a, k_j}) = \frac{exp(\mathbf{a \cdot k_j^T} / \sqrt{d_k})}{\sum_{j'} exp(\mathbf{a \cdot k_{j'}^T} / \sqrt{d_k})}\\
        \mathbf{h_i} &= (\mathbf{v_1, v_2, \cdots, v_i}) \cdot (\alpha_{i1}, \alpha_{i2}, \cdots, \alpha_{ii})^T\\
        \mathbf{y_i} &= \mathbf{h_i \cdot W^O}
    \end{align}

    The design of the anchor vector helps a model to generalize and address the cold start problem. During training, similar questions get similar anchor vector $\mathbf{a}$, therefore helping the model learn weights that generalized to a group of similar questions. During inference, an anchor vector derived from an unseen question represents its similarity to questions in the training set, therefore the model treats it as a mixture of known questions and compute appropriate aggregation weight.

    With this modification, we address the issue of using self-attention on aggregation tasks that the aggregation result should not be oversensitive to the last forecast. Anchor attention not only could be used in forecast aggregation, but it is also generalized able to all sequence summary tasks.
	
	\begin{figure}[h!]
		\centering
		\includegraphics[width=\linewidth]{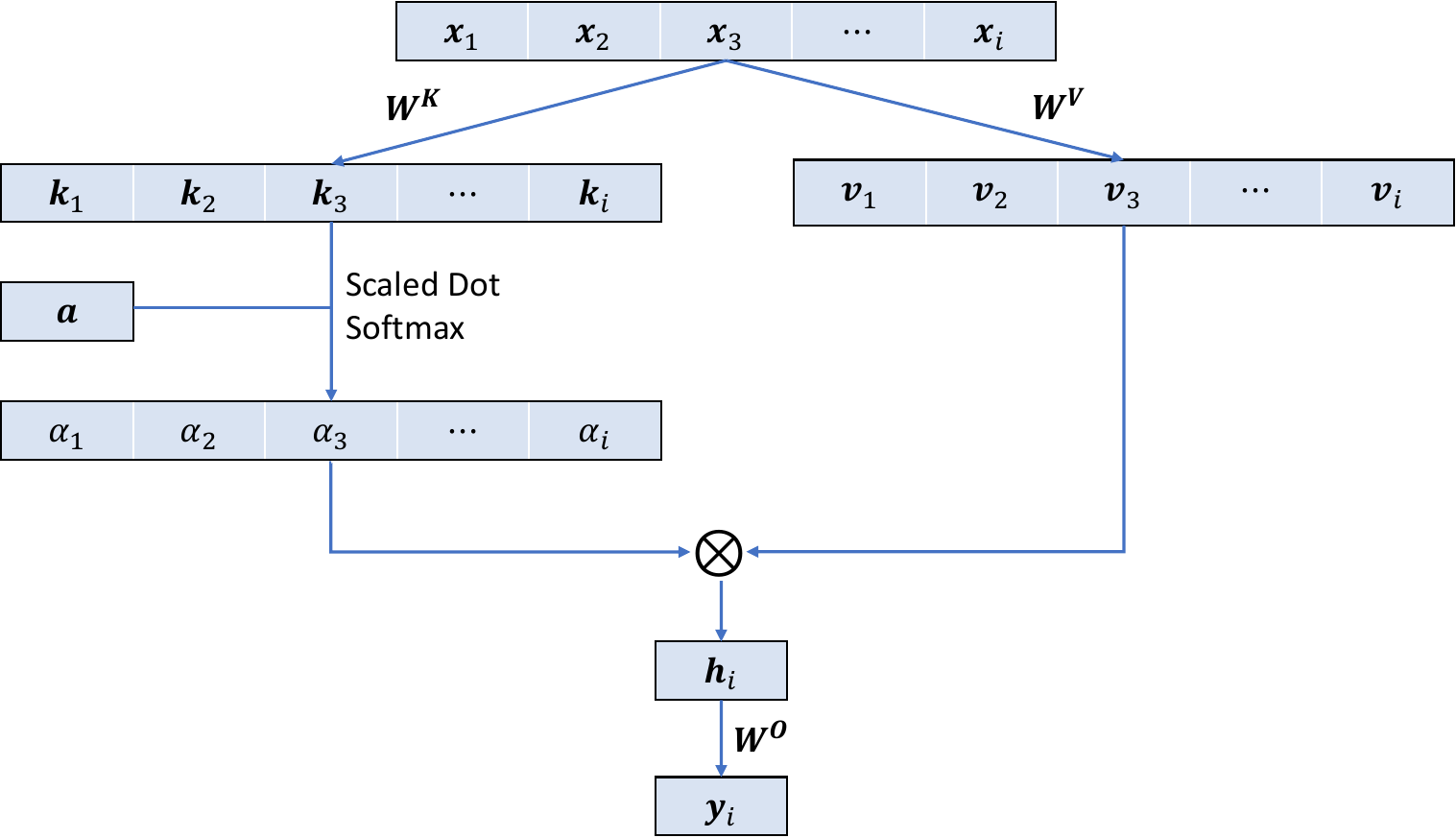}
		\caption{Overview of proposed model.}
		\label{fig:pipeline}
	\end{figure}
	
	\subsection{Model Overview}
	\label{sec:emb_ifp}
	 A graphical overview of the model is presented in Fig. \ref{fig:pipeline}. $X_i$ is the model input and is a concatenation of three parts: $\mathbf{x_i = [f_i, u_i, p_i]}$
	 \begin{enumerate}
	     \item Forecast vector $\mathbf{f_i} \in R^k$, $\sum_j \mathbf{f_{ij}} = 1$
	     
	      For a question with $k$ answer options, we could represent a forecast made by a user as a k-dimension vector $\mathbf{f_i}$, $\mathbf{f_i}$ sum to 1. For instance, a forecast for 5 answer options question could be represented $\mathbf{f_i} = [0, 0.6, 0.4, 0, 0]$. 
	         
	     \item User embedding for that forecaster $\mathbf{u_i}$
	     
	       To identify forecasters and measure their similarity, there is a trainable embedding as part of the model. The embedding is randomly initialized with uniform distribution between $(-\sqrt{3}, \sqrt{3})$ as zero mean, unit variance. $\mathbf{u_i}$ represents user embedding for the author of forecast $\mathbf{x_i}$.

	     \item Position embedding $\mathbf{p_i}$ for location $i$

	     	Although the aggregation results should not be oversensitive to a particular forecast, the temporal order of the forecasts is still valuable, as recent forecasts are usually more accurate information and more valuable. We use sinusoidal position embedding~\cite{Vaswani2017AttentionIA}. It could be viewed as sine and cosine functions of different frequencies. In the equation below, $i$ is the position, $j$ is the dimension, $d$ is the embedding size. Each dimension of the positional encoding corresponds to a sinusoid, and the wavelengths form a geometric progression from $2\pi$ to $10000 \cdot 2\pi$.
	\begin{align}
	    p_{i, 2j} &= sin \left(\frac{i}{10000^\frac{2j}{d}}\right)\\
	    p_{i, 2j+1} &= cos \left(\frac{i}{10000^\frac{2j}{d}}\right)
	\end{align}

	 \end{enumerate}

    The next part is anchor vector $\mathbf{a}$. We use pre-trained word embedding \cite{word2vec} from fastText \cite{fasttext_emb}. For each word $w_i$ in the question text, there is a corresponding vector $\mathbf{v_i} \in \mathbb{R}^{300}$ as its embedding. With bag of words assumption~\cite{bagofword}, the question text embedding $\mathbf{e_{text}}$ could be computed as the average of all word embedding $\mathbf{v_i}$. The anchor vector $\mathbf{a}$ is a projection of $\mathbf{e_{text}}$ with trainable weight $\mathbf{W^A}$.
    
    The aggregated hidden state $\mathbf{h_i}$ is in high dimension space. Weight $\mathbf{W^O}$ is used to project $\mathbf{h_i}$ into k-dimension vector $\mathbf{y_i}$, which sums to 1 and serve as the aggregated forecast.
 
    Our model performs inference at the end of each day, with $\mathbf{x_t}$ being the latest forecast available on that day. Whenever there is a question resolved, our model will be re-trained with the resolved question added into the training set.
	 
    Our methods do not have special treatment for human or machine models. In the rest of the paper, forecasters refer to both human forecasters and machine models.

    \subsection{Toy Example}
    To better illustrate how our model works, we present a toy example. There is a forecast question: \emph{What will be the long-term interest rate for Hungary in July 2018?}. This question has 3 answer options, and the third option is the right answer, as shown in Fig. \ref{fig:example}.  We need to make an aggregated forecast on May 1st, 2018, and there are 10 forecasts currently available, made by 8 forecasters. 
    
    The word embedding table and available forecasts are shown in Fig. \ref{fig:emb}. The word embedding table is pre-trained by fastText\cite{fasttext_emb}, $dim_{word} = 3$. The user embedding is a model weight learned from historical forecasting data, $dim_{user} = 1$. The position embedding is generated according to Eq. (11), $dim_{pos} = 1$.
    
    The list of trained weights is shown in Fig. \ref{fig:weights}. For simplicity, in this toy example, we set $\mathbf{W^A = I_3}, \mathbf{W^O = I_3}$, $\mathbf{W^V}$ as a $5\times3$ zero matrix with top-3 diagonal elements set to 1. $\mathbf{W^K}$ is a model weight learned from training data.
    
    Then question text embedding $\mathbf{e_{\text{text}}}$ could be computed by the average of word embedding.
    Anchor vector $\mathbf{a}$, attention score $\alpha$, output state $\mathbf{h_i}$, aggregated forecast $\mathbf{y_i}$ could be computed according to Eq. (7) (8) (9) (10) respectively. This aggregated forecast having a Brier score $0.1487$.
    \begin{align*}
        \mathbf{e_{\text{text}}}&=[0.1083, 0.1250, 0.1417]\\
        \mathbf{a} & = [0.1079, 0.1244, 0.1407] \\
        \alpha &= [0.0063, 0.0084, 0.0069, 0.0050, 0.0112, 0.0082, 0.4123, 0.0210, 0.5032, 0.0175] \\
        \mathbf{h_i} &= [0.1132, 0.2553, 0.6314] \\
        \mathbf{y_i} &= [0.1132, 0.2553, 0.6314] 
    \end{align*}
    
    This toy example shows several advantages of our proposed method. (1) Forecast \#1 and \#4 are made by the same forecaster, they have the same user embedding, but as their position embedding are different, our model could assign different weights for them (0.0063 vs 0.0050). (2) Our model assigns more weights on accurate forecasts like \#7 and \#9 (3) Our model assigns more weights on recent forecasts, but is still able to identify less accurate forecasts like \#10 and assign less weight.

    In this toy example, we set $\mathbf{W^A, W^O, W^V}$ to the identity matrix, making it looks like a linear model. In actual application, all the weights are learned from training data, and the embedding dimension for word, user, and position would be larger (see Section \ref{sec:config}), making it a deep non-linear aggregation model. Model weights are learned using gradient descent method~\cite{adam} to minimize mean daily Brier Score~\cite{brier1950verification} on available historical forecasting data.

    \begin{figure}[H]
    \settowidth\m{-}
    \settowidth\z{$0$}
    \begin{center}
    \begin{tabular}{ |c|c| } 
     \hline
     Word & Embedding  \\
     \hline
     What & $[\text{-}0.1,\text{-}0.1,\hspace{\m}0.1]$ \\ 
     will & $[\hspace{\m}0.\hspace{\z},\text{-}0.1,\hspace{\m}0.1]$  \\
     be & $[\hspace{\m}0.\hspace{\z},\text{-}0.1,\hspace{\m}0.\hspace{\z}]$  \\ 
     the & $[\text{-}0.1,\hspace{\m}0.\hspace{\z},\hspace{\m}0.\hspace{\z}]$  \\ 
     long-term & $[\hspace{\m}0.5,\hspace{\m}0.9,\hspace{\m}0.1]$  \\ 
     interest & $[\hspace{\m}0.3,\hspace{\m}0.2,\hspace{\m}0.3]$  \\ 
     rate & $[\hspace{\m}0.1,\hspace{\m}0.3,\hspace{\m}0.2]$  \\ 
     for & $[\hspace{\m}0.1,\hspace{\m}0.\hspace{\z},\hspace{\m}0.1]$  \\ 
     Hungary & $[\hspace{\m}0.2,\text{-}0.1,\hspace{\m}0.3]$  \\ 
     in & $[\hspace{\m}0.\hspace{\z},\hspace{\m}0.\hspace{\z},\hspace{\m}0.\hspace{\z}]$  \\ 
     July & $[\hspace{\m}0.\hspace{\z},\text{-}0.1,\hspace{\m}0.5]$  \\ 
     2018 & $[\hspace{\m}0.2,\hspace{\m}0.6,\hspace{\m}0.\hspace{\z}]$  \\ 
     \hline
    \end{tabular}\phantom{X}\begin{tabular}{ |c|c|c|c|c| } 
     \hline
     \# & User Id & Forecast V. & User E. & Position E.\\
     \hline
     1 & 1 & $[0.3,0.4,0.3]$ & $[\hspace{\m}0.33]$ & $[\hspace{\m}0.00]$ \\ 
     2 & 2 & $[0.4,0.4,0.2]$ & $[\hspace{\m}0.24]$ & $[\hspace{\m}0.84]$ \\ 
     3 & 3 & $[0.5,0.4,0.1]$ & $[\hspace{\m}0.27]$ & $[\hspace{\m}0.91]$ \\ 
     4 & 1 & $[0.6,0.2,0.2]$ & $[\hspace{\m}0.33]$ & $[\hspace{\m}0.14]$ \\ 
     5 & 4 & $[0.3,0.3,0.4]$ & $[\hspace{\m}0.04]$ & $[\text{-}0.76]$ \\ 
     6 & 5 & $[0.3,0.4,0.3]$ & $[\hspace{\m}0.11]$ & $[\text{-}0.96]$ \\ 
     7 & 6 & $[0.1,0.3,0.6]$ & $[\text{-}1.36]$ & $[\text{-}0.28]$ \\ 
     8 & 4 & $[0.2,0.2,0.6]$ & $[\hspace{\m}0.04]$ & $[\hspace{\m}0.66]$ \\ 
     9 & 7 & $[0.1,0.2,0.7]$ & $[\text{-}1.23]$ & $[\hspace{\m}0.99]$ \\ 
     10 & 8 & $[0.0,0.6,0.4]$ & $[\hspace{\m}0.05]$ & $[\hspace{\m}0.41]$ \\ 
     \hline
    \end{tabular}
    \end{center}
        \caption{Left: Word embedding table for question text. Right: Table of available forecasts }
        \label{fig:emb}
    \end{figure}

    \begin{figure}[H]
\begin{align*}
W^A = \begin{bmatrix}
    1 & 0 & 0 \\
    0 & 1 & 0 \\
    0 & 0 & 1 \\
    \end{bmatrix}W^K &= \begin{bmatrix}
    \text{-}2.6929 & \text{-}3.6257 & \text{-}3.1801 \\
    \text{-}0.2185 & \text{-}0.4953 & \text{-}0.7536 \\
    \hspace{\m}2.7046 & \hspace{\m}4.8680 & \hspace{\m}4.2125 \\
    \text{-}9.7907 & \text{-}9.9820 & \text{-}11.8808 \\
    \hspace{\m}0.8092 & \hspace{\m}1.5965 & \hspace{\m}1.2388 \\
    \end{bmatrix}
    W^V = \begin{bmatrix}
    1 & 0 & 0 \\
    0 & 1 & 0 \\
    0 & 0 & 1 \\
    0 & 0 & 0 \\
    0 & 0 & 0 \\
    \end{bmatrix}
      W^O = \begin{bmatrix}
    1 & 0 & 0 \\
    0 & 1 & 0 \\
    0 & 0 & 1 \\
    \end{bmatrix}
    \end{align*}
        \caption{Trained Weights}
        \label{fig:weights}
    \end{figure}

	\subsection{Brier Score Loss}
	\label{sec:brier}
	We adopt a loss function that is based on the Brier score~\cite{brier1950verification}, which is widely used to assess the quality of a probabilistic forecast. We will first define the Brier score and then describe how we adapt it as a loss function in our approach.

	\begin{definition}[Unordered Brier Score]
		\label{def:brier}
		Given a forecast $\mathbf{f}=(\mathbf{f_{1}},\mathbf{f_{2}},\ldots,\mathbf{f_{n}})$, and the actual outcome $\mathbf{o}\in \Delta^n$, we use the Brier score $B(\mathbf{f})$ as a measure of accuracy:
		\begin{equation}
		B(\mathbf{f})=\sum_{i=1}^n\left(\mathbf{f_i} -\mathbf{o_i} \right)^2.
		\end{equation}
	\end{definition}
	
	Although the above scoring is only for categorical questions, to address the fact that some questions have ordered outcomes. That is, if the actual outcome is option $k$, choosing option $k-1$ is more favorable to option $k-2$ , we use the ordered Brier score, which is a variant that uses the cumulative probabilities instead of the densities~\cite{jose2009sensitivity}.
	
	\begin{definition}[Ordered Brier Score]
		\label{def:orderd_brier}
		Applying the cumulative sum to the unordered Brier score yields the ordered Brier score.
		\begin{equation}
			B(\mathbf{f})=\frac{1}{n-1}\sum_{i=1}^{n-1}\left[\left(\sum_{j=1}^i \mathbf{f_j} - \sum_{j=1}^i \mathbf{o_j} \right)^2 + \left(\sum_{j=i}^n \mathbf{f_j} - \sum_{j=i}^n \mathbf{o_j} \right)^2\right]
		\end{equation}
	\end{definition}

Mean daily Brier score (MDB), is defined as an unweighted average of the daily Brier score in a question's forecasting period. To evaluate the quality of a forecasting method, mean of mean daily Brier scores (MMDB) is used, which the unweighted average of MDB among all questions. We implemented MMDB as our loss function. 

	The ordered scoring rule is used for ordered categorical questions. Using the ordered Brier score enables the model to learn the ordinal relation between answer options.

	\subsection{Model Configuration Detail}
	\label{sec:config}
	Our model has an adaptive learning rate decay and weight reset mechanism. If the validation set loss continues to increase for 5 epochs, we multiply the current learning rate by 0.95. If the validation set loss keeps increasing for 20 epochs, we reset the network weight to the last state having the lowest validation set loss.
	
	To avoid over-fitting, we utilize dropout~\cite{dropout}, Leaky ReLU~\cite{maas2013rectifier} in our implementation. Other configuration parameters are listed below.

\begin{table}[h!]
\begin{centering}
\caption{Hyper-parameters and training details.}
\begin{tabular}{lc}
\toprule 
Name & Value \tabularnewline
\midrule
Forecast Vector Length & 5 \tabularnewline
User Embedding Dimension & 200 \tabularnewline
Position Embedding Dimension & 95 \tabularnewline
Anchor Vector Length & 300 \tabularnewline
Key, Value Dimension & 64 \tabularnewline
Batch size & 256 \tabularnewline
Training epoch & 100 epochs, with weight reset\tabularnewline
Learning rate & $10^{-4}$, with adaptive decay \tabularnewline
Optimizer & Adam \tabularnewline
\bottomrule 
\label{tab:transformer}
\end{tabular}
\par\end{centering}
\end{table}

	\begin{table}[t]
		\begin{center}
			\caption{Summary statistics of the questions and forecasts in our dataset. STD is the sample standard deviation.}
			\label{tbl:summary}
			%\resizebox{0.5\linewidth}{!}{%
				\begin{tabular}{lccccc}
					\toprule
					Statistic & Dataset & Value\\
					\toprule
					Questions & GJP & 277 \\
					& HFC & 375\\
					Human Forecasters & GJP & 7714\\
					& HFC & 2240\\
					Machine Models & GJP & 0\\
					& HFC & 3\\
					\toprule
					Statistic & Dataset & Min & Median &
					Mean (STD) & Max\\
					\midrule
					Forecasts per Question & GJP & 76 & 2,101 & 2,946.8 (2,531.7) & 19,578\\
					 & HFC & 9 & 192 & 265.2 (204.6) & 1,029\\
					Forecasts per User & GJP & 1 & 14 & 30.3 (40.3) & 277\\
					 & HFC & 1 & 32 & 44.4 (63.2) & 1,339\\
					Users per Question & GJP & 69 & 540 & 844.8 (709.5) & 4,255\\
					 & HFC & 8 & 108 & 173.6 (148.9) & 669\\
					Days Question is Open & GJP & 2 & 81 & 99.7 (72.9) & 285\\
					 & HFC & 1 & 42 & 23.8 (42.2) & 184\\
					\bottomrule
				\end{tabular}
			
			\par\end{center}
	
	\end{table}
	\section{Experimental Results}
	We utilized human and machine forecasts on a wide range of forecasting problems from our hybrid competition dataset and good judgment open dataset in our evaluation. Some statistics of these two datasets are listed in Table~\ref{tbl:summary}.
	
	\subsection{Baseline Methods}\label{sec:baseline_methods}
	We compare our approach to \cite{atanasov2017distilling,l2reg}, who proposed an aggregation method using temporal decay, differential weighting based on past performance, and extremization. The approach outlined by \cite{atanasov2017distilling} provides a simple yet strong performance baseline that outperforms prediction markets for distilling the wisdom of crowds.
	The approach is the weighted aggregation of individual predictions for each question into a single forecast:
	$$
	\overline{\mathbf{f}}_{t,q}\propto\sum_{t,i}d_{t}\times w_{t,i}^\gamma\times \mathbf{f}_{t,i,q},
	$$
	where, $\overline{\mathbf{f}}_{t,q}$ is the weighted aggregate forecast for question $q$ across all individual forecasts. $\mathbf{f_{t,i,q}}$ coming from user $i$ at time $t$ for question $q$; $d_{t}$ is a temporal decay parameter which gives more weight to recent forecasts; $w_{t,i}$ are individual weights for each forecaster $i$ and time $t$ based on their past performance; and $\gamma$ is the exponent parameter for the individual weights.
	Finally, the aggregate forecast $\overline{\mathbf{f}}_{t}$ is extremized based on Karmarkar's equation~\cite{karmarkar1978subjectively} by a linear transformation in
	the log-odds scale to move the distribution away from the uniform distribution $1/a$, as follows:
	$$
	\log\left(\frac{(a-1)\hat{\mathbf{f}}_{t,k}}{1-\hat{\mathbf{f}}_{t,k}^{\alpha}}\right)=\alpha\left(\frac{(a-1)\overline{\mathbf{f}}_{t,k}}{1-\overline{\mathbf{f}}_{t,k}}\right),
	$$
	where $\hat{\mathbf{f}}_{t}$ is the final forecast probability distribution; $\alpha$ is the extremization parameter; $a$ is the number of answer options, such that $k \in \{1, 2, 3, \cdots, a \}$.
	For ordinal questions, we use the cumulative probabilities instead of the densities.
	
	We consider three variants based on \cite{atanasov2017distilling}:
	\begin{description}
		\item[M0:] Unweighted average with temporal decay.
		\item[M1:] Weighted average with temporal decay and extremization.
		\item[M2:] Weighted average with temporal decay, differential weighting based on past performance and extremization.
	\end{description}

	All parameters were learned from the training data: accuracy weights for each participant were set to the inverse of their average Brier across the questions they answered; the rest of the parameters were estimated using grid search.

	We also include a heuristic approach, \textbf{Top Individuals}. After the first 10 questions are resolved, the top 40\% of forecasters are chosen and labeled “Top Individuals.” This list of top individuals is updated after the conclusion of each question to maintain the top 40\% of forecasters by Brier score.

	\begin{table}[h!]
		\begin{center}
			\caption{Our reproduced Top Individuals Methods}
			\begin{tabular}{ lcccc }
				\toprule
				Method & Mean Brier (Published) & Mean Brier (Reproduced)\\
				\midrule
				Top Indivs Mean & 0.2319 & 0.2159\\
				Top Indivs Median & 0.1632 & 0.1947\\
				\bottomrule
			\end{tabular}
		\end{center}
		\label{tab:repoduced}
	\end{table}
	
	For the GJP dataset, we have reproduced the Top Individuals method following the setup described the published paper~\cite{l2reg}. We are calculating the mean daily Brier score over the period 2013-07-08 to 2015-06-09. A comparison between published scores and reproduced results is shown in Table 3. Our reproduced scores are slightly different from the published score, as mentioned in the paper there are 1,799 forecasters in the final year, while the released dataset has 6858 forecasters in the final year. In the rest of the paper, we will use our reproduced results as a baseline to compare with other methods.
	
	\subsection{Brier Score Comparison}
    \begin{table}[h!]
    \vspace{1em}
	\begin{tabular}{lcc}
    Dataset: HFC\\
    \toprule
    Method & Mean & STD \\
    \midrule
	M0 & 0.3208  & 0.2742 \\
	M1 & 0.3186  & 0.2743 \\
	M2 & 0.3041   & 0.3210 \\
	Self-Attention & 0.2865  & 0.4057 \\
	\textbf{Anchor-Attention} & \textbf{0.2515}  & 0.2655 \\
	\bottomrule
    \end{tabular}
    \quad
    \begin{tabular}{lcc}
    Dataset: GJP\\
	\toprule
	Method & Mean &  STD \\
	\midrule
	Top Indivs Mean & 0.2159 & 0.0539 \\
	Top Indivs Median & 0.1947 & 0.0754 \\
	Self-Attention & 0.1804 &  0.1221 \\
	\textbf{Anchor-Attention} & \textbf{0.1211}  & 0.0474\\
	\bottomrule
    \end{tabular}
    
    \caption{Comparison of Brier score across different methods}
    \label{tab:scores}
    \end{table}
    
	We evaluate our model on both our hybrid forecasting dataset and GJP dataset. We are comparing the mean and variance of mean daily Brier scores (MDB) for all forecasting questions in each dataset. We simulate a real setup that re-train our model every time a question is closed.
	%We perform 5-fold cross-validation. Cross-validation is performed at the question level, meaning that all forecasts for a particular question are constrained within their respective fold. We record the mean and standard deviation of the Brier score across folds, and the quantile range of all Brier scores in Table~\ref{tab:scores}. Our model has the lowest average Brier score and quantile scores, along with the lowest variance, suggesting that our model is very stable.
	
	For our hybrid forecasting dataset, we performed a one-sided test between M2 (strongest baseline) and our Anchor Attention method. The null hypothesis is that the population mean Brier score of the Anchor Attention method is not lower than the M2 method. There are $N=375$ questions in the dataset. According to Table \ref{tab:scores}, we could compute test statistic is $z=\frac{\bar{X_1}-\bar{X_2}}{\sqrt{\frac{\sigma_1^2}{N_1}+\frac{\sigma_2^2}{N_2}}}=\frac{0.3041-0.2515}{\sqrt{\frac{0.3209^2}{375}+\frac{0.2655^2}{375}}}=2.445$ which translates to $p = 0.0072$. This p-value rejects the null hypothesis at $\alpha=0.01$. We also perform similar test on GJP dataset ($N=277$) between Top Indivs Median and Anchor-Attention method, test statistic is $z = \frac{0.1947-0.1211}{\sqrt{\frac{0.0754^2}{277}+\frac{0.0474^2}{277}}}=13.75$, the corresponding p value is less than $10^{-5}$. This p-value also rejects the null hypothesis at $\alpha=0.01$
	
	Both tests conclude that our Anchor Attention-based method gets a lower Brier score than M2 methods and Top Indivs Median, the improvement is significant.

	\subsubsection{Analysis of the distribution of Brier scores}
	\begin{figure}[h]
		\centering
		\includegraphics[width= 0.49\linewidth]{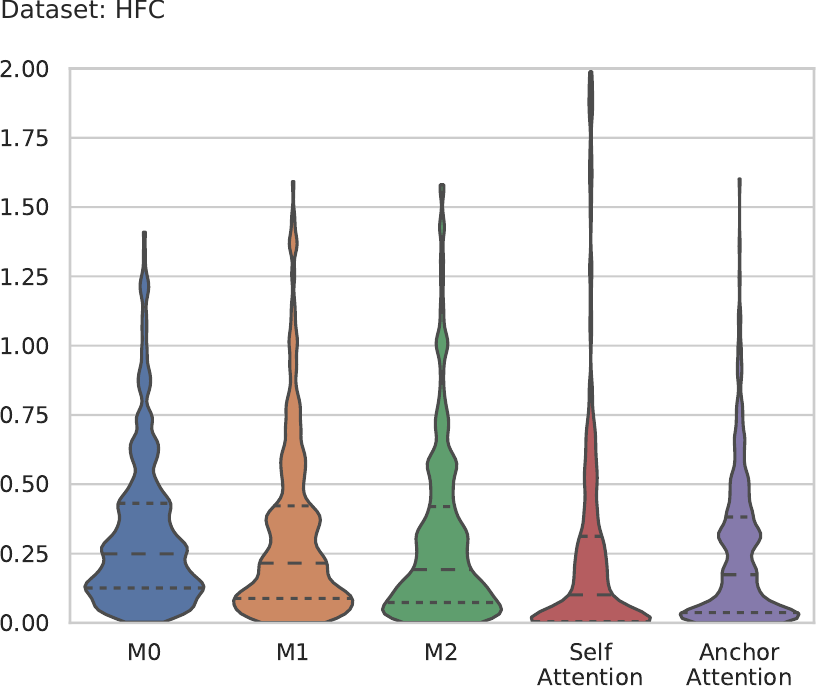}
		\includegraphics[width= 0.49\linewidth]{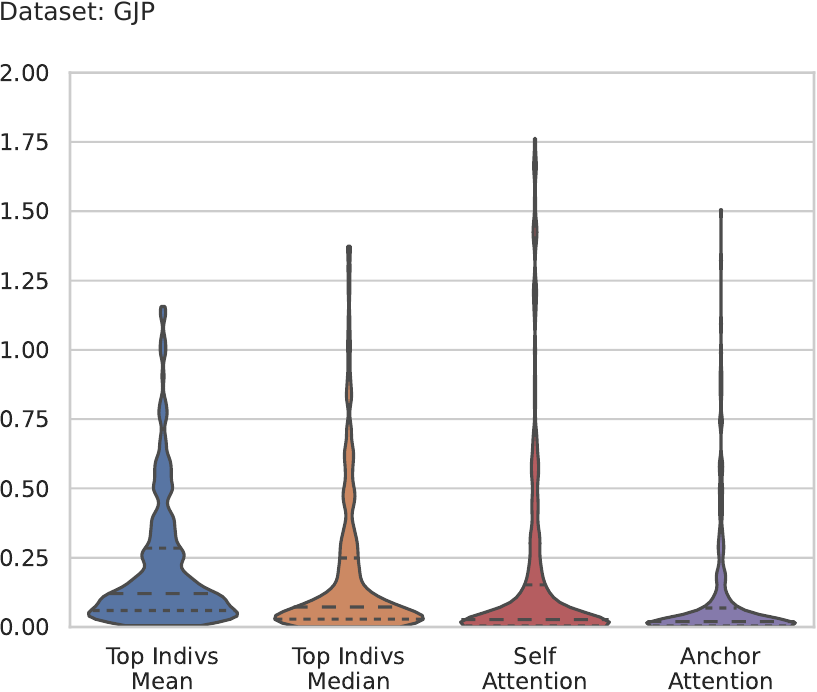}
		\caption{The distribution of the Brier scores for each method.}
		\label{fig:brierdist}
	\end{figure}
	
	Figure~\ref{fig:brierdist} shows a violin plot depicting the distribution of Brier scores of each method. We evaluate the Brier score distribution of aggregated forecasts. The width of each violin is proportional to the ratio of forecasts having that Brier score. The broken lines represent 25\%, 50\%, and 75\% quantile.
	We observed that our proposed method achieves the lowest average Brier scores and quantile scores. The highest (worst) Brier score is also comparable to the baseline model, which means our model is robust.

	Appendix Figures~\ref{fig:calibration} and \ref{fig:roc} provide alternative ways, independent of Brier, to compare the methods based on calibration and discrimination (AUC and ROC Curve) which are common in classification tasks. The Anchor Attention method has better discrimination with an AUC of 0.9, while the baseline methods have 0.85.

	\subsubsection{Percentile of aggregated forecast among all individual forecasters}
	Next, we use the quantile metric~\cite{han2019universal} to compare the performance of each ensemble method by its relative position on the cumulative distribution of the corresponding individual scores. We rank each aggregated forecast among all the individual forecasts. The percentile of the ranking is plotted.
	
\begin{figure}
		\centering
		\includegraphics[width=0.49\linewidth]{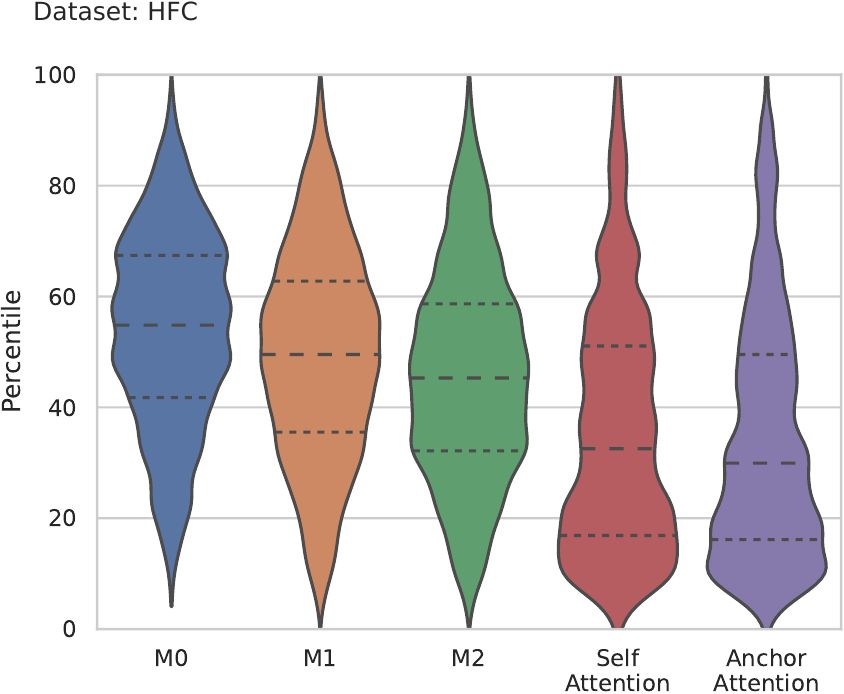}
		\includegraphics[width=0.49\linewidth]{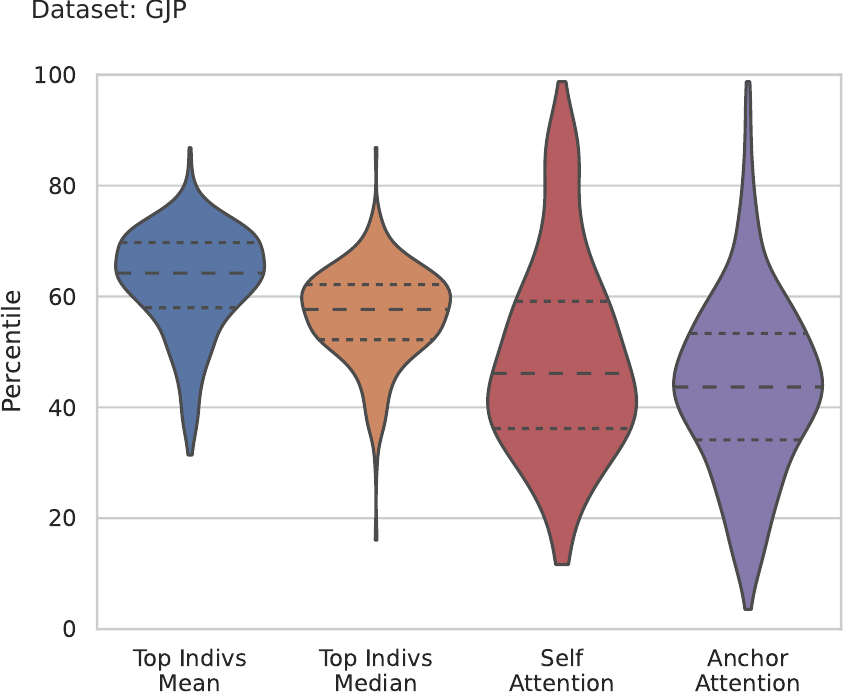}
		\caption{The distribution of ranks. A method that has more forecasts in lower positions is better than one that has more forecasts in higher ones. }
		\label{fig:rankdist}
	\end{figure}

	In Figure~\ref{fig:rankdist} we plot the distribution of the ranking of aggregated forecasts among all individual forecasts. The smaller the ranking, the lower the Brier score. 
	We observe that for the Anchor Attention model, the aggregated forecasts are better than most individual forecasts. For example, half of the time, our model beats 70\% of the individual forecasts, while baseline methods beat approximately 50\% of the individual forecasts. 

	\subsubsection{Performance in forecasting period}
	
\begin{figure}[h!]
    \centering
   \includegraphics[width=0.49\linewidth]{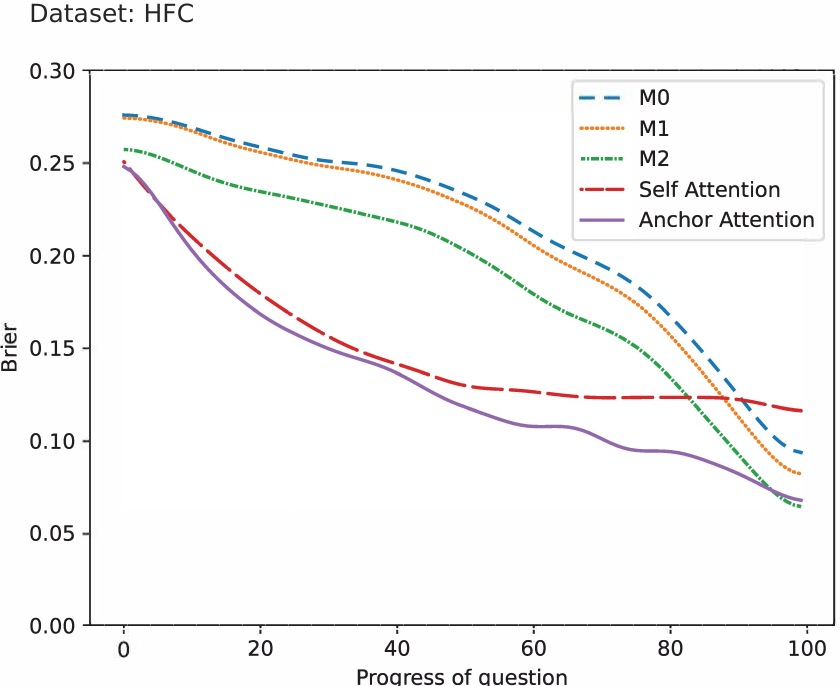}
   \includegraphics[width=0.49\linewidth]{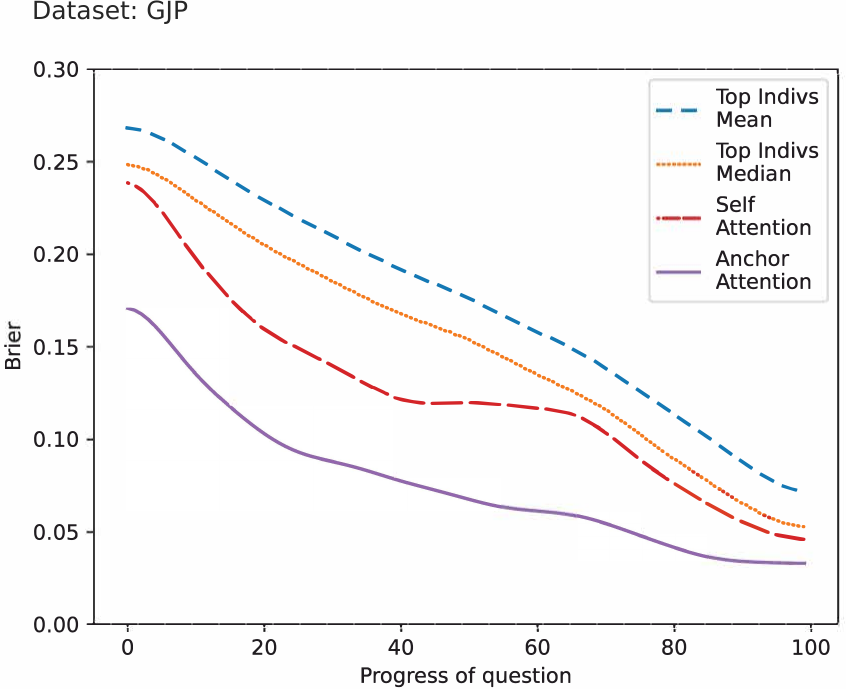}
    \caption{Average Brier score across all questions along the forecast progress}
    \label{fig:brier_time}
\end{figure}

	 We also compare the performance of our methods as a function of how much time a question has been open, shown in Figure~\ref{fig:brier_time}. We normalize the progress in the forecast period as 0 to 1 and calculate the average Brier score across all questions along with the forecast progress. We use 100 bins, and the results are smoothed with Gaussian kernel, sigma = 5.

	As expected, the closer to the resolution date, the better \emph{all} aggregation methods perform. However, we observed that our Anchor attention model quickly reduce its Brier score at the early stages of the question, and its overall Brier score is also lower. The insight is that our model can learn to find similarities between questions and forecasters, so when a new question arrives, it can quickly find the right forecasters to trust. 

    \subsection{Visualization of Question Embedding}

\begin{figure}[h]
    \centering
     \includegraphics[width=0.8\linewidth]{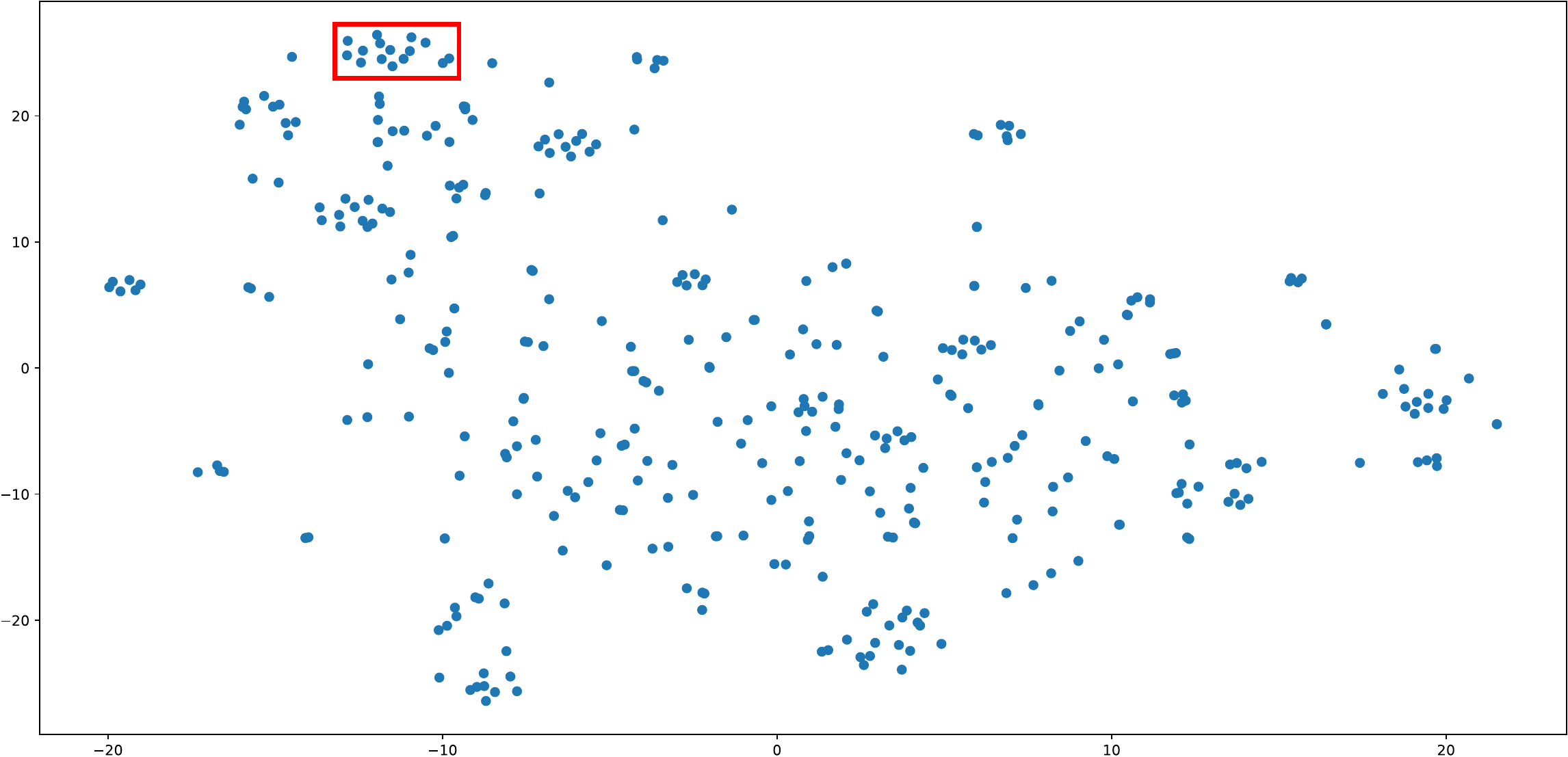}
     
     \vspace{0.5em}
     \includegraphics[width=0.8\linewidth]{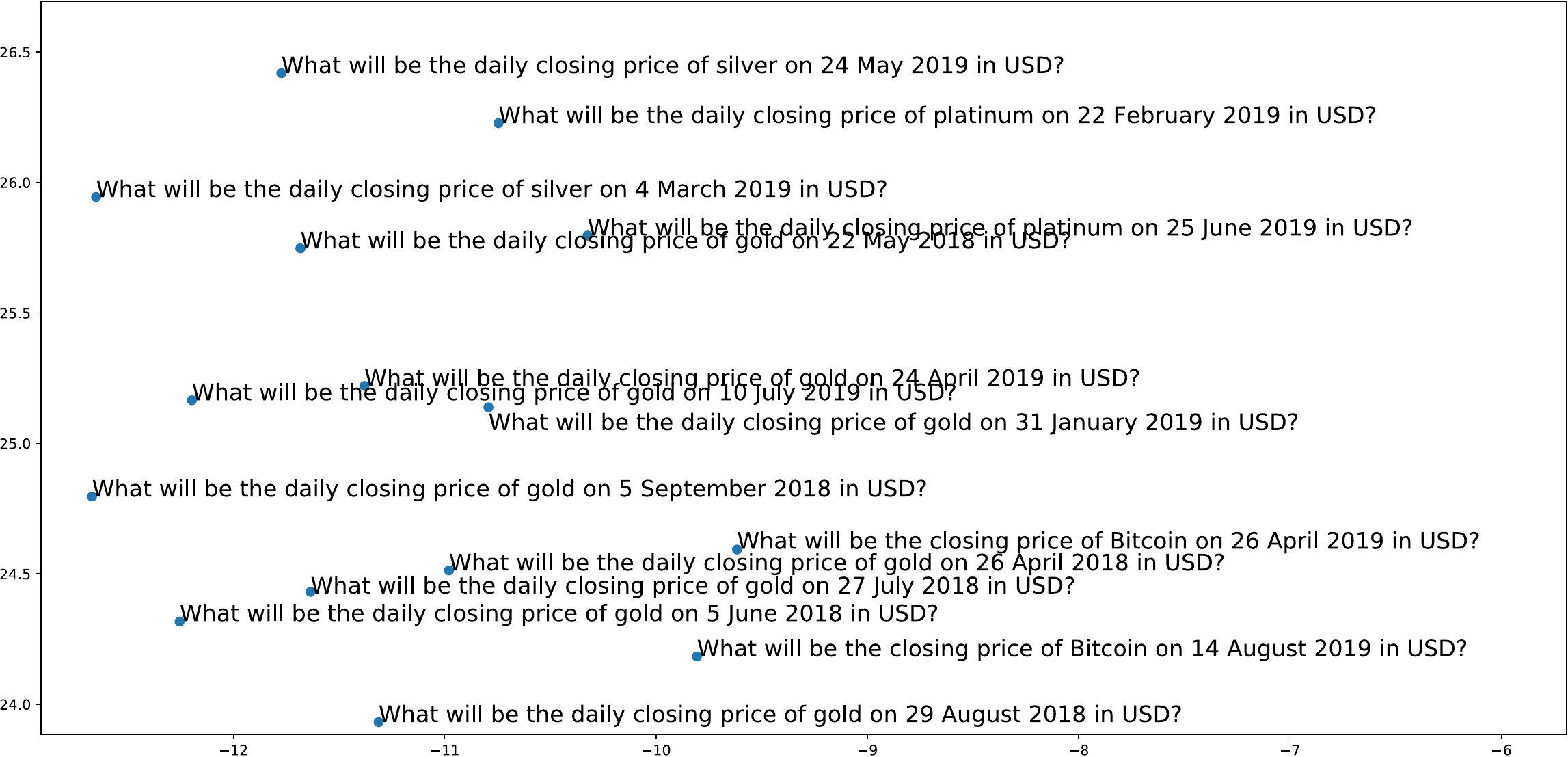}
    \caption{Question Embedding with t-SNE dimension reduction}
    \label{fig:ifp_emb}
    \centering
\end{figure}
    From Figure \ref{fig:ifp_emb}, we find that questions are not uniformly distributed in the embedding space; instead, they form various clusters. So we had the lower figure zoom in to the red rectangle region in the upper figure. We find those closely-located questions are similar; they are all asking about the closing price of precious metal on different dates. It demonstrates that our model learned a good representation of questions, and is aware of the similarity between questions.

	\subsection{Visualization of User Embedding}
	\begin{figure}[h]
    \centering
	\includegraphics[width=0.8\linewidth]{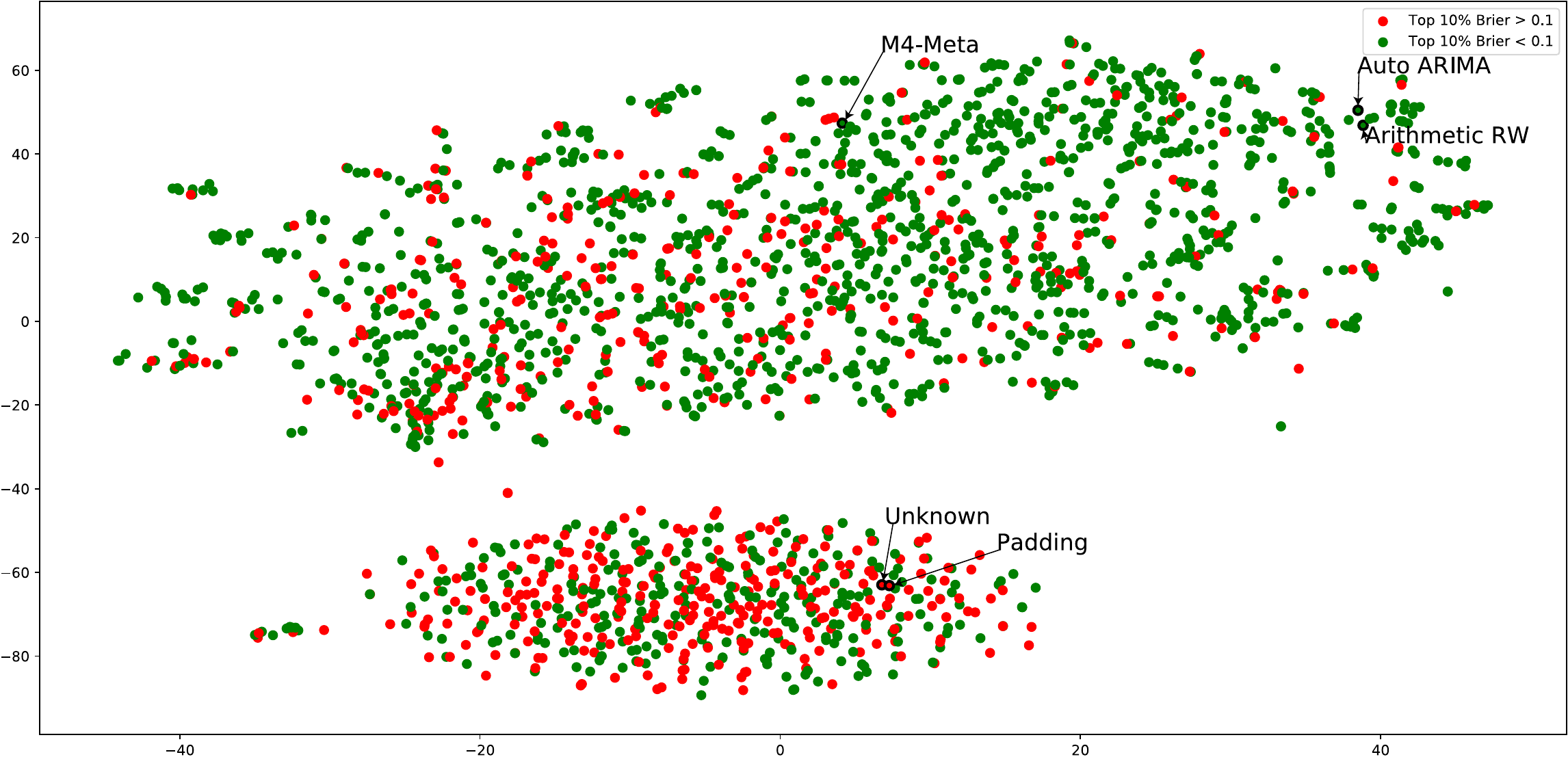}
    \caption{User embedding with t-SNE dimension reduction}
    \label{fig:user_emb}
\end{figure}
    We find the forecasters are also not uniformly distributed; they roughly formed two clusters. Interestingly, two machine models, Auto ARIMA and Arithmetic RW, are closely located in the embedding space, while they are intrinsically similar (Arithmetic RW is a special case of Auto ARIMA model). If we define good forecasters as those whose top 10\% Brier score is smaller than 0.1, we find the good forecasters are mostly located in the upper cluster. It demonstrates that our model learned a good representation of forecasters.
	
	\subsection{Analysis of Attention Scores}
\begin{figure}[t]
		\centering
		\includegraphics[width=0.8\linewidth]{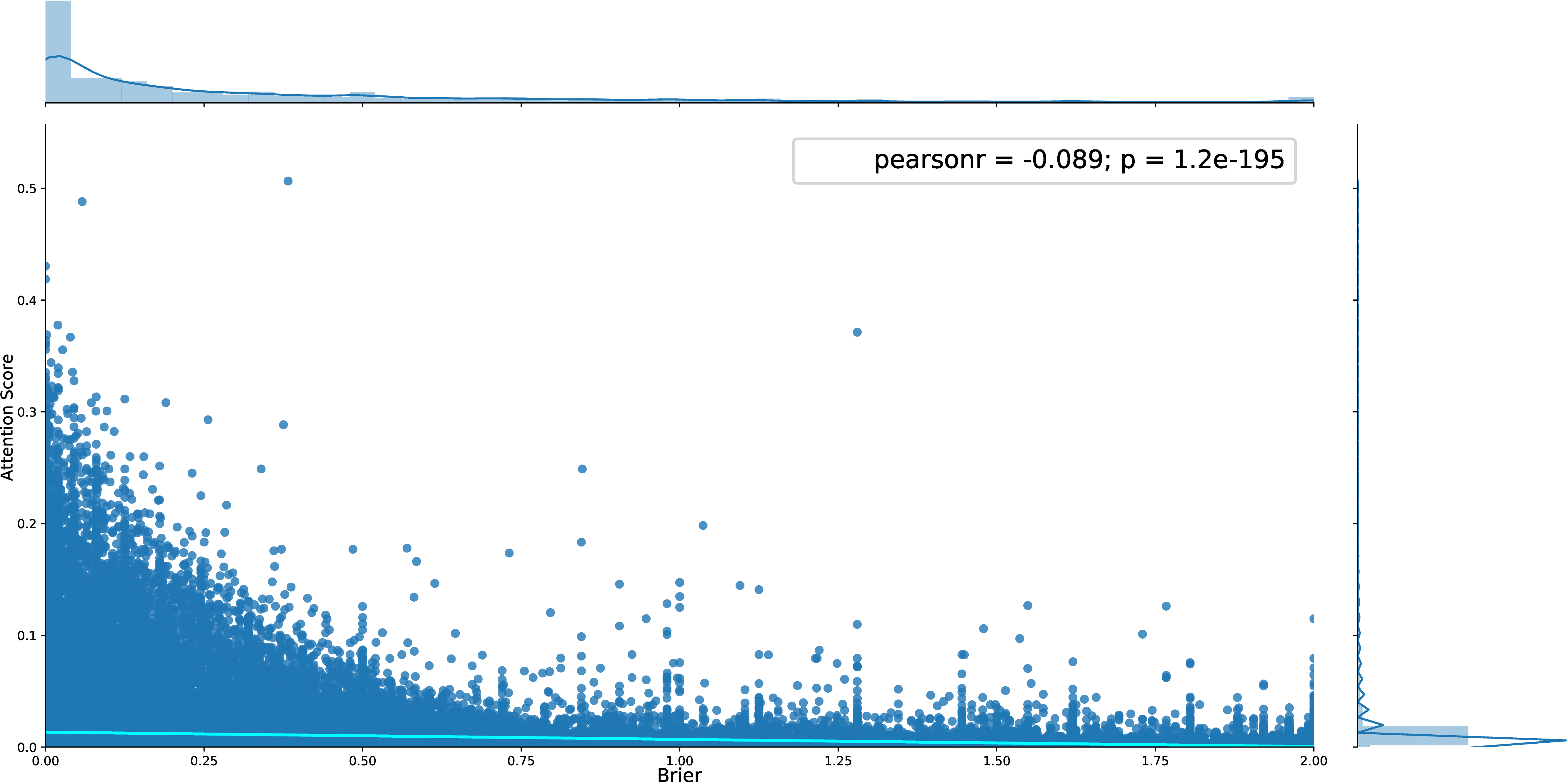}
		\caption{Relation between attention score to Brier score. The trend line shows a downward trend, indicating that higher attention scores are given to forecasts with lower Brier scores.}
		\label{fig:attention}
	\end{figure}

	Figure~\ref{fig:attention} shows forecasts' attention scores against their Brier scores.
	We observe that the attention score is negatively (coef=\text{-}0.089, p=1.2e-195) correlated to the Brier score, which means good forecasts (low Brier) have higher weights in aggregation. Also, we observe there are no points in the upper right corner, which means forecasts with a high Brier score do not receive a high attention score. This suggests that our model is learning representations that are useful to distinguish good forecasts from bad ones. 

	\section{Conclusion}
	We propose a novel approach to aggregate a hybrid set of forecasts based upon deep neural networks. Using data from real-world forecasting competitions, we demonstrate that this approach outperforms the current state-of-the-art in forecast aggregation. Moreover, we showed preliminary evidence that through learning an embedding of questions and forecasters, our approach learns to identify more accurate forecasters while simultaneously combining human forecasters with machine models (See Appendix Table~\ref{sup:ablation}).
	
	Although this work is focused on identifying quality forecasters in a geopolitical event forecasting context, the results here can be applied in other settings. For example, this approach can be used to identify which workers are most appropriate to choose for a given task in a micro-tasking environment. Future work will aim to better understand the implications of this method in other environments and to adapt our approach to identifying quality workers with a minimal amount of data.

%\section*{Acknowledgements}
%This research is based upon work supported in part by the
%Office of the Director of National Intelligence (ODNI), Intelligence Advanced Research Projects Activity (IARPA), via
%2017-17071900005. The views and conclusions contained
%herein are those of the authors and should not be interpreted
%as necessarily representing the official policies, either expressed or implied, of ODNI, IARPA, or the U.S. Government. The U.S. Government is authorized to reproduce and
%distribute reprints for governmental purposes notwithstanding any copyright annotation therein.

\appendix
\section{Ablation Study}
This section is devoted to answering the following questions
\begin{enumerate}
    \item Does identifying the past performance of each forecaster help the aggregation?
    \item Does use both human and machine forecasts help the aggregation?
\end{enumerate}
We restrict the analyses to the subset of questions that have both machine and human forecasts available (even in configurations that only use humans or machine models). This subset is different from the set of questions used in previous experiments. We conducted a 2x3 factorial design experiment. The first factor is whether or not we can identify (with user embedding) the forecaster of each forecast. When user embedding is removed, the model can not learn each forecaster's past performance and give personalized aggregation weights. The second factor is the type of forecasts used as input to the aggregation: only human forecasts, only machine forecasts, and both.

Figure~\ref{sup:ablation} compares the Brier scores for this experiment. First, the model's performances are improved with the addition of user embedding, suggesting that the model learns to accurately weigh forecasters based on their past performance. Second, with humans only or machine models only, the score is not as good as with both; however, this is only true if user embedding is used. With these findings, we conclude the two questions we raised earlier both turned out to be true, that user embedding and using a hybrid approach is beneficial to forecast aggregation.

It's interesting to see human forecasters are more accurate than machine forecasters, one reason could be the time series used in machine models are also available for human forecasters to analyze and interpret, while human forecasters could integrate additional information to make better forecasts.

\begin{table}[h!]
\caption{Performance results for different inputs.}
\label{sup:ablation}
\begin{center}
\begin{tabular}{ |c|c|c| } 
\hline
User Embedding & Forecaster type & Brier score  \\
\hline
No & Human & 0.303 \\
\hline
No & Machine & 0.319 \\
\hline
No & Human + Machine & 0.303 \\
\hline
Yes & Human & 0.297 \\
\hline
Yes & Machine & 0.318 \\
\hline
Yes & Human + Machine & \textbf{0.290}\\
\hline
\end{tabular}
\end{center}
\end{table}

\section{Calibration and Discrimination}
\begin{figure}[h!]
    \centering
   \includegraphics[width=0.8\linewidth]{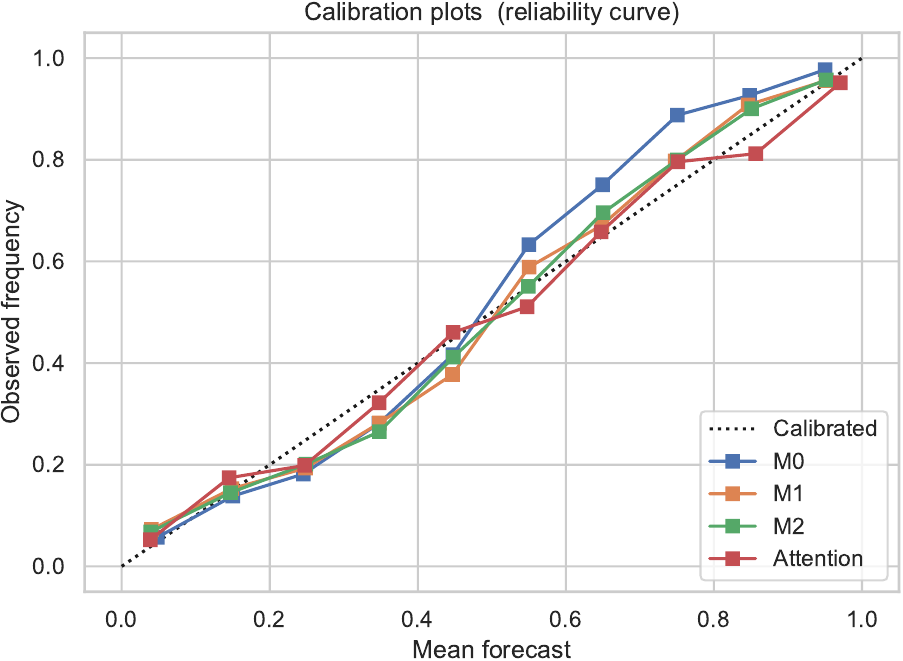}
    \caption{Calibration Plots }
    \label{fig:calibration}
\end{figure}

Calibration refers to the ability to make forecasts that coincide with the observed empirical frequencies of events being predicted. Calibration error is zero when the events with predicted probabilities p actually occur about p percentage of the time. Therefore, any deviation from the diagonal line suggests poor calibration. Figure~\ref{fig:calibration} shows the calibration curves for each method. The x-axis is divided into 10 bins, each spanning 10\% on the probability scale. The y-axis denotes the observed proportion of events that fall in each bin.

\begin{figure}[h!]
    \centering
   \includegraphics[width=0.8\linewidth]{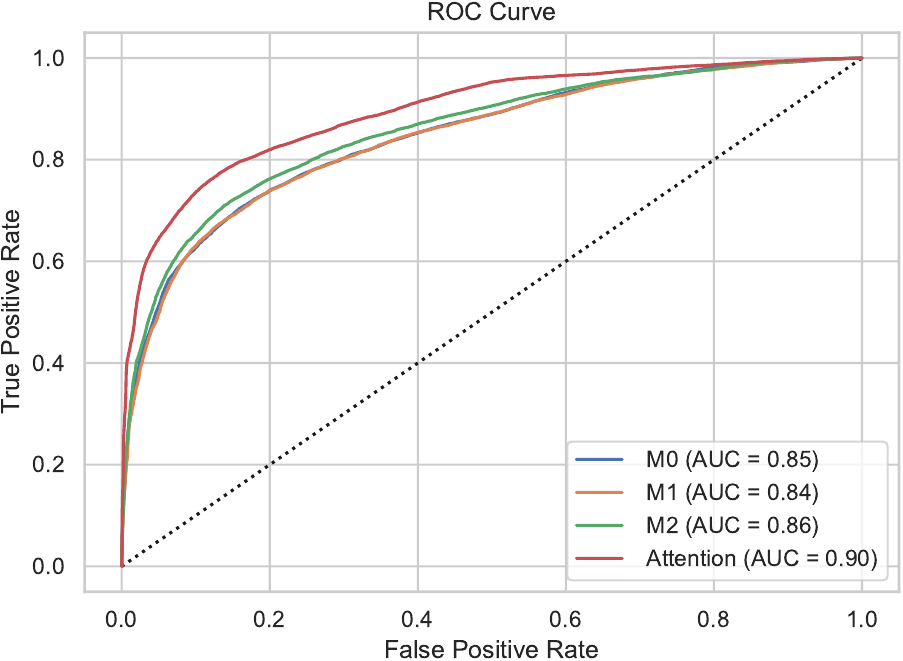}
    \caption{ ROC Curves  }
    \label{fig:roc}
\end{figure}

Discrimination refers to the capacity to distinguish between true positive outcomes and true negative ones.
We use the AUC (Area Under The Curve) ROC (Receiver Operating Characteristics) curve to evaluate the discrimination of the models. The ROC curve plots the probability of a true positive against that of a false positive. Perfect resolution means all the probability mass is under the curve. Figure~\ref{fig:roc} shows the ROC curves for all methods.

\bibliography{elsarticle-template}

\begin{thebibliography}{10}
\expandafter\ifx\csname url\endcsname\relax
  \def\url#1{\texttt{#1}}\fi
\expandafter\ifx\csname urlprefix\endcsname\relax\def\urlprefix{URL }\fi
\expandafter\ifx\csname href\endcsname\relax
  \def\href#1#2{#2} \def\path#1{#1}\fi

\bibitem{atanasov2017distilling}
P.~Atanasov, P.~Rescober, E.~Stone, S.~A. Swift, E.~Servan-Schreiber,
  P.~Tetlock, L.~Ungar, B.~Mellers, Distilling the wisdom of crowds: Prediction
  markets vs. prediction polls, Management science 63~(3) (2017) 691--706.

\bibitem{l2reg}
D.~Cross, J.~Ramos, B.~Mellers, P.~E. Tetlock, D.~W. Scott, {Robust forecast
  aggregation: Fourier L2E regression}, Journal of Forecasting 37~(3) (2018)
  259--268.
\newblock \href {http://dx.doi.org/10.1002/for.2489}
  {\path{doi:10.1002/for.2489}}.

\bibitem{GJP}
G.~J. Project, \href{https://doi.org/10.7910/DVN/BPCDH5}{{GJP Data}} (2016).
\newblock \href {http://dx.doi.org/10.7910/DVN/BPCDH5}
  {\path{doi:10.7910/DVN/BPCDH5}}.
\newline\urlprefix\url{https://doi.org/10.7910/DVN/BPCDH5}

\bibitem{tetlock2017expert}
P.~E. Tetlock, Expert political judgment: How good is it? How can we know?,
  Princeton University Press, 2017.

\bibitem{tetlock2016superforecasting}
P.~E. Tetlock, D.~Gardner, Superforecasting: The art and science of prediction,
  Random House, 2016.

\bibitem{satopaa2014combining}
V.~A. Satop{\"a}{\"a}, J.~Baron, D.~P. Foster, B.~A. Mellers, P.~E. Tetlock,
  L.~H. Ungar, Combining multiple probability predictions using a simple logit
  model, International Journal of Forecasting 30~(2) (2014) 344--356.

\bibitem{Vaswani2017AttentionIA}
A.~Vaswani, N.~Shazeer, N.~Parmar, J.~Uszkoreit, L.~Jones, A.~N. Gomez,
  L.~Kaiser, I.~Polosukhin, Attention is all you need, in: NIPS, 2017.

\bibitem{armstrong1989combining}
J.~S. Armstrong, Combining forecasts: The end of the beginning or the beginning
  of the end?, International Journal of Forecasting 5~(4) (1989) 585.

\bibitem{clemen1989combining}
R.~T. Clemen, Combining forecasts: A review and annotated bibliography,
  International journal of forecasting 5~(4) (1989) 559--583.

\bibitem{jose2008simple}
V.~R.~R. Jose, R.~L. Winkler, Simple robust averages of forecasts: Some
  empirical results, International journal of forecasting 24~(1) (2008)
  163--169.

\bibitem{winkler2019probability}
R.~L. Winkler, Y.~Grushka-Cockayne, K.~C. Lichtendahl~Jr, V.~R.~R. Jose,
  Probability forecasts and their combination: A research perspective, Decision
  Analysis 16~(4) (2019) 239--260.

\bibitem{lichtendahl2020some}
K.~C. Lichtendahl~Jr, R.~L. Winkler, Why do some combinations perform better
  than others?, International Journal of Forecasting 36~(1) (2020) 142--149.

\bibitem{brown1991forecast}
L.~D. Brown, Forecast selection when all forecasts are not equally recent,
  International Journal of Forecasting 7~(3) (1991) 349--356.

\bibitem{ranjan2010combining}
R.~Ranjan, T.~Gneiting, Combining probability forecasts, Journal of the Royal
  Statistical Society: Series B (Statistical Methodology) 72~(1) (2010) 71--91.

\bibitem{baron2014extremizing}
J.~Baron, B.~Mellers, P.~Tetlock, E.~Stone, L.~Ungar, Distilling the wisdom of
  crowds: Prediction markets vs. prediction polls, Decision analysis 11~(2)
  (2014) 133--145.

\bibitem{satopaa2014probability}
V.~A. Satop{\"a}{\"a}, S.~T. Jensen, B.~A. Mellers, P.~E. Tetlock, L.~H. Ungar,
  et~al., Probability aggregation in time-series: Dynamic hierarchical modeling
  of sparse expert beliefs, The Annals of Applied Statistics 8~(2) (2014)
  1256--1280.

\bibitem{gaunt2016training}
A.~Gaunt, D.~Borsa, Y.~Bachrach, Training deep neural nets to aggregate
  crowdsourced responses, in: Proceedings of the Thirty-Second Conference on
  Uncertainty in Artificial Intelligence, AUAI Press, 2016, p. 242251.

\bibitem{crowd_ranking}
G.~Nebbione, D.~Doran, S.~Nadella, B.~Minnery, Deep neural ranking for
  crowdsourced geopolitical event forecasting, in: {\'E}.~Renault,
  P.~M{\"u}hlethaler, S.~Boumerdassi (Eds.), Machine Learning for Networking,
  Springer International Publishing, Cham, 2019, pp. 257--269.

\bibitem{NIPS2014_a14ac55a}
I.~Sutskever, O.~Vinyals, Q.~V. Le, Sequence to sequence learning with neural
  networks, in: Z.~Ghahramani, M.~Welling, C.~Cortes, N.~Lawrence, K.~Q.
  Weinberger (Eds.), Advances in Neural Information Processing Systems,
  Vol.~27, Curran Associates, Inc., 2014.

\bibitem{Zhang2019SequencetoSequenceAM}
J.-X. Zhang, Z.~Ling, L.~Liu, Y.~Jiang, L.~Dai, Sequence-to-sequence acoustic
  modeling for voice conversion, IEEE/ACM Transactions on Audio, Speech, and
  Language Processing 27 (2019) 631--644.

\bibitem{10.1145/3419106}
T.~Shi, Y.~Keneshloo, N.~Ramakrishnan, C.~K. Reddy, Neural abstractive text
  summarization with sequence-to-sequence models, ACM/IMS Trans. Data Sci.
  2~(1).

\bibitem{Elman90findingstructure}
J.~L. Elman, Finding structure in time, COGNITIVE SCIENCE 14~(2) (1990)
  179--211.

\bibitem{JORDAN1997471}
M.~I. Jordan, Chapter 25 - serial order: A parallel distributed processing
  approach, in: J.~W. Donahoe, V.~{Packard Dorsel} (Eds.), Neural-Network
  Models of Cognition, Vol. 121 of Advances in Psychology, North-Holland, 1997,
  pp. 471--495.

\bibitem{lstm}
S.~Hochreiter, J.~Schmidhuber, Long short-term memory, Neural Comput. 9~(8)
  (1997) 1735--1780.

\bibitem{wolfers2004prediction}
J.~Wolfers, E.~Zitzewitz, Prediction markets, Journal of economic perspectives
  18~(2) (2004) 107--126.

\bibitem{yeh2006using}
P.~F. Yeh, Using prediction markets to enhance us intelligence capabilities,
  Studies in Intelligence 50~(4) (2006) 137--149.

\bibitem{mellers2015identifying}
B.~Mellers, E.~Stone, T.~Murray, A.~Minster, N.~Rohrbaugh, M.~Bishop, E.~Chen,
  J.~Baker, Y.~Hou, M.~Horowitz, et~al., Identifying and cultivating
  superforecasters as a method of improving probabilistic predictions,
  Perspectives on Psychological Science 10~(3) (2015) 267--281.

\bibitem{brier1950verification}
G.~W. Brier, Verification of forecasts expressed in terms of probability,
  Monthly weather review 78~(1) (1950) 1--3.

\bibitem{ungar2012the}
H.~L. Ungar, B.~Mellers, V.~Satopää, P.~Tetlock, J.~Baron, The good judgment
  project: A large scale test of different methods of combining expert
  predictions, AAAI Fall Symposium: Machine Aggregation of Human Judgment.

\bibitem{arima}
R.~Hyndman, Y.~Khandakar, Automatic time series forecasting: The forecast
  package for r, Journal of Statistical Software, Articles 27~(3) (2008) 1--22.

\bibitem{m4}
S.~Makridakis, E.~Spiliotis, V.~Assimakopoulos, The m4 competition: 100,000
  time series and 61 forecasting methods, International Journal of Forecasting.

\bibitem{joint_align}
D.~Bahdanau, K.~Cho, Y.~Bengio, \href{http://arxiv.org/abs/1409.0473}{Neural
  machine translation by jointly learning to align and translate}, in:
  Y.~Bengio, Y.~LeCun (Eds.), 3rd International Conference on Learning
  Representations, {ICLR} 2015, San Diego, CA, USA, May 7-9, 2015, Conference
  Track Proceedings, 2015.
\newline\urlprefix\url{http://arxiv.org/abs/1409.0473}

\bibitem{gru}
K.~Cho, B.~van Merri{\"e}nboer, C.~Gulcehre, D.~Bahdanau, F.~Bougares,
  H.~Schwenk, Y.~Bengio,
  \href{https://www.aclweb.org/anthology/D14-1179}{Learning phrase
  representations using {RNN} encoder{--}decoder for statistical machine
  translation}, in: In Proc. Conference on Empirical Methods in Natural
  Language Processing ({EMNLP}), 2014, pp. 1724--1734.
\newblock \href {http://dx.doi.org/10.3115/v1/D14-1179}
  {\path{doi:10.3115/v1/D14-1179}}.
\newline\urlprefix\url{https://www.aclweb.org/anthology/D14-1179}

\bibitem{elstm}
Y.~Su, C.-C.~J. Kuo,
  \href{http://www.sciencedirect.com/science/article/pii/S0925231219306289}{On
  extended long short-term memory and dependent bidirectional recurrent neural
  network}, Neurocomputing 356 (2019) 151 -- 161.
\newblock \href
  {http://dx.doi.org/https://doi.org/10.1016/j.neucom.2019.04.044}
  {\path{doi:https://doi.org/10.1016/j.neucom.2019.04.044}}.
\newline\urlprefix\url{http://www.sciencedirect.com/science/article/pii/S0925231219306289}

\bibitem{bert}
J.~Devlin, M.-W. Chang, K.~Lee, K.~Toutanova,
  \href{https://aclanthology.org/N19-1423}{{BERT}: Pre-training of deep
  bidirectional transformers for language understanding}, in: Proceedings of
  the 2019 Conference of the North {A}merican Chapter of the Association for
  Computational Linguistics: Human Language Technologies, Volume 1 (Long and
  Short Papers), Association for Computational Linguistics, Minneapolis,
  Minnesota, 2019, pp. 4171--4186.
\newblock \href {http://dx.doi.org/10.18653/v1/N19-1423}
  {\path{doi:10.18653/v1/N19-1423}}.
\newline\urlprefix\url{https://aclanthology.org/N19-1423}

\bibitem{gpt}
A.~Radford, J.~Wu, R.~Child, D.~Luan, D.~Amodei, I.~Sutskever, Language models
  are unsupervised multitask learners.

\bibitem{t5}
A.~Roberts, C.~Raffel, K.~Lee, M.~Matena, N.~Shazeer, P.~J. Liu, S.~Narang,
  W.~Li, Y.~Zhou, Exploring the limits of transfer learning with a unified
  text-to-text transformer, Tech. rep., Google (2019).

\bibitem{word2vec}
T.~Mikolov, I.~Sutskever, K.~Chen, G.~Corrado, J.~Dean, Distributed
  representations of words and phrases and their compositionality, in:
  Proceedings of the 26th International Conference on Neural Information
  Processing Systems - Volume 2, NIPS'13, Curran Associates Inc., USA, 2013,
  pp. 3111--3119.

\bibitem{fasttext_emb}
T.~Mikolov, E.~Grave, P.~Bojanowski, C.~Puhrsch, A.~Joulin, Advances in
  pre-training distributed word representations, in: Proceedings of the
  International Conference on Language Resources and Evaluation (LREC 2018),
  2018.

\bibitem{bagofword}
Z.~Harris,
  \href{https://link.springer.com/chapter/10.1007/978-94-009-8467-7_1}{Distributional
  structure}, Word 10~(2-3) (1954) 146--162.
\newblock \href {http://dx.doi.org/10.1007/978-94-009-8467-7_1}
  {\path{doi:10.1007/978-94-009-8467-7_1}}.
\newline\urlprefix\url{https://link.springer.com/chapter/10.1007/978-94-009-8467-7_1}

\bibitem{adam}
D.~P. Kingma, J.~Ba, \href{http://arxiv.org/abs/1412.6980}{Adam: {A} method for
  stochastic optimization}, in: Y.~Bengio, Y.~LeCun (Eds.), 3rd International
  Conference on Learning Representations, {ICLR} 2015, San Diego, CA, USA, May
  7-9, 2015, Conference Track Proceedings, 2015.
\newline\urlprefix\url{http://arxiv.org/abs/1412.6980}

\bibitem{jose2009sensitivity}
V.~R.~R. Jose, R.~F. Nau, R.~L. Winkler, Sensitivity to distance and baseline
  distributions in forecast evaluation, Management Science 55~(4) (2009)
  582--590.

\bibitem{dropout}
N.~Srivastava, G.~Hinton, A.~Krizhevsky, I.~Sutskever, R.~Salakhutdinov,
  Dropout: A simple way to prevent neural networks from overfitting, Journal of
  Machine Learning Research 15 (2014) 1929--1958.

\bibitem{maas2013rectifier}
A.~L. Maas, A.~Y. Hannun, A.~Y. Ng, Rectifier nonlinearities improve neural
  network acoustic models, in: Proc. icml, Vol.~30, 2013, p.~3.

\bibitem{karmarkar1978subjectively}
U.~S. Karmarkar, Subjectively weighted utility: A descriptive extension of the
  expected utility model, Organizational behavior and human performance 21~(1)
  (1978) 61--72.

\bibitem{han2019universal}
Y.~Han, D.~Budescu, A universal method for evaluating the quality of
  aggregators, Judgment and Decision Making 14~(4) (2019) 395.

\end{thebibliography}

\end{document}